\documentclass[aps,prb,floatfix,groupedaddress]{revtex4}
\usepackage{epsfig}
\usepackage[american]{babel}
\usepackage{amsmath}
\usepackage[dvips]{color}
\usepackage[latin2]{inputenc}
\usepackage[T1]{fontenc}
\usepackage[american]{babel}
\usepackage{color}
\usepackage{graphicx}
\usepackage{graphics}

\newcommand{\LD}{D}
\newcommand{\GD}{\Sigma}
\newcommand{\grgd}{\operatorname{d}}
\newcommand{\hc}{\mbox{$h. c.$}}

\newcommand{\IM}{\operatorname{Im}}
\newcommand{\RE}{\operatorname{Re}}
\newcommand{\sign}{\operatorname{sign}}

\usepackage[bookmarks=true]{hyperref}

\usepackage[ps2pdf]{thumbpdf} 

\begin{document}


\title{Locally gauge invariant theory of large $U_d$ high-T$_c$ cuprates}

\author{S. Bari\v si\' c}

\email{sbarisic@phy.hr}

\affiliation{Department of Physics, Faculty of Science, University of Zagreb, Bijeni\v cka c. 32, HR-10000 Zagreb, Croatia}

\author{O. S. Bari\v si\' c}

\affiliation{Institute of Physics, Bijeni\v cka c. 46, HR-10000 Zagreb, Croatia}
\affiliation{Jo\v zef Stefan Institute, SI-1000 Ljubljana, Slovenia}

\begin{abstract}

The large $U_d$ theory is constructed for the metallic state of high-T$_c$ cuprates. It is based on the Emery three band model extended with the O$_x$-O$_y$ hopping $t_{pp}$ in the $U_d\rightarrow\infty$ limit. The $U_d\rightarrow\infty$ mapping on the slave fermion theory is used. The time-dependent diagrammatic theory in terms of the Cu-O hopping $t_{pd}$ starts from the locally gauge invariant nondegenerate unperturbed state with vanishing average occupation $n_d^{(0)}$ of the Cu state and builds a finite $n_d$ in higher orders. This theory is locally gauge invariant asymptotically, but replaces the $d-p$ anticommutation of the fermions on the Cu and O sites by the commutation and is antisymmetrized {\it a posteriori}. Rather than $t_{pd}$, the small parameter of the theory is $n_d\leq1/2$. The lowest order of the $U_d=\infty$ theory generates the $U_d=0$ single particle Dyson propagators of the hybridized $pdp$- and $dpd$-fermions which exhibit the covalent three band structure filled up to the appropriate chemical potential $\mu$. The leading many body effect is band narrowing, different from that found in mean-field slave boson theories. It is accompanied by the broad incoherent background related to the dynamical quantum charge-transfer disorder associated to the d$^{10}$$\leftrightarrow$d$^9$ Cu/O$_2$ intracell charge transfer fluctuations. The disorder effects fall well below the Fermi level and break the Luttinger sum rule for the conduction band. Those results and even the infinite order non-crossing approximation are insensitive to the omission of the $d-p$ anticommutation rules. The contributions affected by the $d-p$ commutation show up in single particle propagators beyond the third order. The effective local repulsion between the hybridized $pdp$ propagators turns out to be a sizeable $t_{pd}^4/\Delta_{d\mu}^3$, where $\Delta_{d\mu}$ is the difference between the energy of the Cu-site and $\mu$. The {\it a posteriori} antisymmetrization of the theory removes the triplet repulsion between the $pdp$ particles but keeps the singlet repulsion which favors at low energy the incommensurate SDW correlations. Such $t_{pd}^4/\Delta_{d\mu}^3$  repulsion is the metallic counterpart of the $U_d=\infty$ super exchange $J_{pd}$ between the $dpd$ propagators. Resonant valence bonds appear thus as incoherent perturbative corrections here. The resulting modified slave fermion theory (MSFT) approximately obeys the local gauge invariance and conserves the local fermionic anticommutation rules, provided that $n_d$ is sufficiently small. The corresponding theoretical predictions compare favorably with ARPES, NQR, X-ray, neutron scattering, Raman, optic and superconductivity measurements, emphasizing the importance of oxygen degrees of freedom in the physics of high-T$_c$ cuprates.

\end{abstract}

\maketitle

\pagenumbering{arabic}

\section{Introduction\label{Sec01}}

The long standing question in high-T$_c$ cuprates concerns the nature of interactions which are responsible for the superconductivity and the other unusual properties of those materials. The early high-energy spectroscopic measurements indicated that the Hubbard interaction $U_d$ on the Cu site might be quite large. This opened the question of whether or not $U_d$ alone can account for the basic physics of the high-T$_c$ cuprates. Such a question can be rephrased\cite{fr1,an1} in terms of the structure of the effective interactions in high-T$_c$ cuprates, which include the concomitant strong electronic correlations. This fundamental question is discussed in some detail here from the theoretical point of view, with the results finally confronted to some salient experiments. 

The observed phase diagram of cuprates is characterized by a crossover (ignoring the small interplanar couplings) between the insulating long or short range AF phase at small hole doping $0<x<x_{cs}$ to the disordered metallic/SDW/superconducting phase for $x>x_{cs}$. Typical experimental values of $x_{cs}$ found from ARPES\cite{in1,zu1} are of the order of a few percent. The local properties related to the measured ARPES spectra are the average single-particle occupations of the Cu and O$_{x,y}$ sites $n_d$ and $n_p$. The latter can be found in the $x>x_{cs}$ metallic phase\cite{ku1} from the electric field gradients measured\cite{hr1,tg1} by NQR, giving  $n_d$ around $1/2$ which increases on doping with holes. While an accurate evaluation of $n_d$ in the $x<x_{cs}$ regime is hindered by strong local magnetic fields in the AF phase and the narrow range of its stability, it is usually inferred\cite{mt1} that $n_d$ decreases slightly on doping. It is important to note that the NQR results in the well-developed metallic phase rely only on the measured local symmetry of the average charge distribution in the vicinity of the Cu and O nuclei. They are thus essentially model independent.

The crossover is also clearly evidenced by transport,\cite{sz1,wa1,bo1,ma1} optical\cite{uc1} and Raman\cite{gz1,ts1} measurements. In addition to the low frequency conductivity, which presents an unconventional behavior for $x>x_{cs}$, the strong optic edge, associated for $x<x_{cs}$ with the excitation through a gap, is smeared out\cite{uc1} for $x>x_{cs}$ into the transitions between broad structures with finite threshold frequencies. Consistently, the Raman data exhibit\cite{si2, bl1} a neat two magnon AF resonance for $x<x_{cs}$ that smoothens into a broad continuum for $x>x_{cs}$. This observation directly reveals a deconstruction of the AF order upon doping.

There are many other features of cuprates which also corroborate the proposed crossover picture. One such example is the incommensurate magnetic ordering. This ordering is associated with the magnetic superlattice Bragg spots at $\vec q_{SDW}$, which differs\cite{tr2,tr3,st1,ki1,gz1,ya1,vt1} from $\vec q_{AF}=\vec G/2$, where $\vec G=2\pi[1,1]/a$ is the reciprocal lattice wave vector ($a$ is the Cu-Cu lattice constant). Furthermore, symmetry analysis shows that the incommensurate magnetic order may be accompanied by a $\vec q_0$ modulated charge transfer (CT's) within the CuO$_2$ unit cell and the CDW among the cells, coupled in turn linearly to the incommensurate lattice deformations. The deformations give rise to the nuclear superlattice\cite{bi1,sg1,tr3,gz1} Bragg spots, with small wave vector $\vec q_0$. Experimentally, $q_{SDW}$ and $\vec q_0$ are related through the simple Umklapp relation\cite{ki1} $\vec q_0+\vec G=2\vec q_{SDW}$, irrespective of the value of $x$. When only one leg $\vec q_{SDW}$ (and $\vec q_0$) of the wave vector star is present the D$_4$ symmetry is broken and the entangled\cite{bb1} magnetic, CT, CDW charge transfer and lattice deformations appear\cite{bi1,sg1} in what is often called nematic stripes.\cite{ki1,vt1} For $x<x_{cs}$, $\vec q_{SDW}$ (and $\vec q_0$) lie along the diagonals of the CuO$_2$ zone (diagonal stripes) but rotate\cite{ts1} by $\pi/4$ for $x>x_{cs}$ to the positions along the main CuO$_2$ axes (collinear stripes). They show a {\it commensurate} LTO/LTT instability\cite{ax1,kmr,ft1} which gives rise to the new Bragg spot at $2\pi[1,0]/a$ that is {\it coexisting} with collinear stripes. According to ARPES data, the LTO/LTT lattice instability occurs for $x\approx1/8$ close to the doping\cite{in1,yo1,vl1} $x_{vH}>x_c$ which puts the Fermi level on the van Hove singularity of the conduction band. This relates a prominent band feature, namely the $vH$ singularity, to a phase transition observed for $x>x_{cs}$. Remarkably, the LTO/LTT lattice instability suppresses\cite{ax1} the superconductivity in LBCO that would otherwise be close to optimal. In addition, the spin and charge disorders are observed in the glassy and metallic phases of cuprates by many experimental methods, including NMR,\cite{al1,go3,jl1} transport,\cite{bo1,ma1} IR\cite{sz1,wa1} and Raman spectroscopy.\cite{cp1,gz1} 

The simultaneous appearance of magnetic and lattice Bragg spots is a clear signature of the spin and charge coherence in the ground state of the system. The superconductivity, which occurs exclusively\cite{bo1,ma1,ts1} for $x>x_{cs}$, is itself a coherent state. Apparently, those coherent features are not only balanced among themselves but also compete with the spin and charge disorders. The main aim of the present work is to contribute to the understanding of these competitions.

\section{Weak versus strong coupling}

The theories of high-T$_c$ cuprates often start from the tight binding model, with the vacuum consisting of Cu$^+$(d$^{10}$) and O$^{2-}$(p$^6$) states. The Cu$^{2+}$(d$^9$) and O$^-$(p$^5$) site energies are denoted then by $\varepsilon_d$, $\varepsilon_{px}$ and $\varepsilon_{py}$ with $\Delta_{pd}=\varepsilon_p-\varepsilon_d>0$ in the hole language and $\Delta_{pp}=\varepsilon_{px}-\varepsilon_{py}$. The Cu$^{3+}$(d$^8$) state is reached by spending the energy $2\varepsilon_d+U_d$ where $U_d$ describes the bare interaction of two holes on the Cu site, which may be reduced to some extent\cite{fr1} by intra atomic correlations. In contrast to that the O(p$^4$) configuration is usually associated with the energy $2\varepsilon_p$ i.e. $U_p$ is considered as relatively small. Referring\cite{em1} to LDA results, such a model of the Cu and O sites was completed with the Cu-O hybridization $t_{pd}$. This selection of the relevant single particle and interaction parameters is often called the Emery model.\cite{em1} The original Emery model was later extended\cite{km1} by the direct O-O hopping $t_{pp}$ which describes the hole propagation rotated by $\pi/4$ with respect to the CuO$_2$ axes. The model is completed by choosing the total number of holes $1+x$ per CuO$_2$ unit cell, where $x$ is the average number of doped carriers. The average single particle occupations of the Cu and O$_{x,y}$ sites $n_d$ and $n_p$ are then linked by the sum rule $n_d+2n_p=1+x$.

In order to explain the coherent features of cuprates, some early theoretical works\cite{fr1,bs5} invoked a small $U_d$ regime of the $t_{pp}=0$ Emery model. This was later extended,\cite{fr3} under the assumption $U_d<\Delta_{pd}$, to an effective intraband $U_d$  reduced by metallic kinematic\cite{ka1} correlations. The $x=0$ Fermi surface touches the logarithmic vH singularities at T=0, irrespective of the ratio $t_{pd}/\Delta_{pd}$. The equal sharing of charge $n_d=1/2$ at $x=0$ between Cu and two O's is obtained in the covalent limit\cite{fr1,fr3,bs5} $t_{pd}\gg\Delta_{pd}$, whereas $n_d=1$ corresponds to the opposite ionic limit $\Delta_{pd}\gg t_{pd}$. The $x=0$ Fermi surface is perfectly nested.\cite{fr1,dz2,bs5} Finite $U_d$, and especially its Umklapp component,\cite{bs5} enhances therefore quite strongly the commensurate $\vec q_{AF}=\vec G/2$ SDW fluctuations. The latter may give rise\cite{ru1} to the unconventional behavior of the conductivity. The $\vec q=0$ O$_x$/O$_y$ CT fluctuations within the CuO$_2$ unit cell are also enhanced, whether coupled\cite{bb1} or not\cite{bs2,bb1} to the $\vec G/2$ SDW. The $\vec q=0$ Cu/O$_2$ and O$_x$/O$_y$ CT´s make the Raman active\cite{ku3} quadrupole moment of the CuO$_2$ unit cell vary, while conserving\cite{bs2} its total charge. Therefore, in contrast to the optically active $\vec q\rightarrow0$ CDW\cite{ru1,ku2,ku3} or Raman active A$_{1g}$ Cu-O CT,\cite{ku3} the $\vec q\rightarrow0$ B$_{1g}$ or B$_{2g}$ intracell charge (or current) fluctuations are not frustrated by the long-range Coulomb forces.\cite{ku3} When coupled linearly to the acoustic modes, the O$_x$/O$_y$ CT may induce the acoustic lattice instability\cite{bb1} at $\vec q=0$. The $\vec q=0$ O$_x$/O$_y$ CT also couples quadratically\cite{bs2} to the very slow tilting modes in lanthanum cuprates which results\cite{bs2,pu1} in the LTO/LTT instability of LBCO. With finite doping $x$ the SDW instability moves\cite{sc1,sk3} to incommensurate values of $\vec q_{SDW}$ and becomes weaker in this model. In particular the effect of the $\vec G$-Umklapp interaction $U_d$ in the build up of AF correlations is diminished in this way. Usually, the Umklapp in question is removed by hand from the theory when $\vec q_{SDW}$ becomes incommensurate, i.e. the SDW commensurability pinning together with the resulting intrinsic striping and disorder is ignored. This results in a (too) smooth sliding of $\vec q_{SDW}$ with $x$. The incommensurate O$_x$/O$_y$ CT fluctuations show similar behavior, assuming that they are driven by two SDW's, which gives rise\cite{ki1,bb1,vt1} to the observed relation $\vec q_0+\vec G=2\vec q_{SDW}$ between the incommensurate lattice and magnetic wave vectors. Finally, in the presence of the attractive interactions between carriers, the weak-coupling prediction is\cite{fr1,dz2,bs5} that the SDW instability is replaced by the superconductivity either by doping $x$ or at $x=0$ by (chemical) pressure. Although the SDW behavior emerges more or less correctly, the main problem of this description is that it puts\cite{bs2} the commensurate LTO/LTT instability at $x=0$ rather than at sizeable hole doping $x\approx x_{vH}\approx1/8$, where it is observed\cite{ax1,ft1,tr3} in LBCO. 

This could be remedied by including $t_{pp}$. Although smaller than $t_{pd}$ on chemical grounds, it is quite relevant in the weak coupling theory. First, for $t_{pp}<0$ (appropriate\cite{km1,mr1} for high T$_c$ cuprates), $t_{pp}$ sets\cite{km1,bs2,mr1} the Fermi level of the $x=0$ half filled lowest band below the vH singularity, which means that the latter is reached upon a finite hole doping $x_{vH}>0$. The ARPES spectra of the hole-doped cuprates in the $x>x_{cs}$ metallic state can than be fit\cite{km1,mr1} by the Emery three-band structure. Those fits indicate\cite{mr1} that the bare parameters obey the relation $\Delta_{pd}^2\gtrsim2t^2_{pd}\gtrsim\Delta_{pd}|t_{pp}|\gtrsim4t_{pp}^2$ with $|t_{pp}|$ large enough to account\cite{km1} for the $\pi/4$ rotation of the Fermi surface (Fermi arcs) with respect to its $t_{pp}=0$ form. Concomitantly, the Cu occupation $n_d$ at $x=0$ is reduced below $1/2$ in the covalent limit $t_{pd}>\Delta_{pd}$. Furthermore, $t_{pp}$ breaks\cite{xu1,qs1} the perfect nesting properties of the $x=0$ Fermi surface, i.e., using the 1d language,\cite{bb1} it plays the role of the imperfect nesting parameter. The elementary SDW particle-hole bubble develops then the peaks at incommensurate $\vec q_{SDW}$ for $x=x_{vH}$ and small $\omega$.\cite{xu1,km1,sk3} When the small interaction $U_d$ is introduced, the resulting $\vec G$-Umklapp scattering of two particles is in discord with this value of the wave vector and the resulting value of $\vec q_{SDW}=(\vec q_0+\vec G)/2$ is, in general, incommensurate and weakly affected by a small $U_d$. On the other hand, the elementary O$_x$/O$_y$ CT particle-hole bubble for $x=x_{vH}$ is\cite{bb1} logarithmically singular at $\omega=0$, $\vec q=0$ for any value of $t_{pp}$, corresponding to the Jahn-Teller splitting\cite{fr1,bs2,pu1} of the vH singularities, and favoring the commensurate LTT instability. While this latter result agrees with observations\cite{ax1,ft1,tr3} in lanthanates, the problem is that in the weak coupling theory, strong magnetic correlations occur only for $x\approx x_{vH}$. As mentioned above, this particular feature is at variance with observations where appreciable magnetic correlations coupled to the lattice (stripes) persist over a wide range of doping, from $x>0$ at least up to optimal dopings. While the weak coupling theory with finite $t_{pp}$ explains thus the metallic phase reasonably well, it fails to describe the Mott-AF phase for $x<x_{cs}$.

Another important feature not encountered in conventional weak coupling theories is the intrinsic CT disorder d$^{10}+$p$^4$$\leftrightarrow$d$^9+$p$^5$ (p$^n$'s will be dropped from now on), as well as the corresponding d$^9$ spin disorder on Cu-sites. Already in the earliest theoretical works with $U_d\gg\Delta_{pd}$, it was pointed out\cite{go1} that the static d$^{10}$$\leftrightarrow$d$^9$ disorder is an essential feature of the $t_{pd}=0$ limit when $\varepsilon_d$  falls within the dispersive band, $\Delta_{pd}\leq4|t_{pp}|$ in the present language. A finite $t_{pd}$ is expected to render the d$^{10}$$\leftrightarrow$d$^9$ disorder dynamic. Indeed, a broad dynamic background appeared in the early\cite{ni1} slave particle NCA calculation and also in the non-magnetic\cite{zl1} and magnetic\cite{mc1,rb1} DMFT calculations with $U_d\geq\Delta_{pd}$. However the relation between the broad background and the dynamic d$^{10}$$\leftrightarrow$d$^9$ disorder had not been established until recently.\cite{bb1} It is noteworthy that the strong ionic electron-phonon\cite{go1,bs2} coupling may make the d$^{10}$$\leftrightarrow$d$^9$ disorder slow again by a polaronic reduction\cite{go1,go3} of single particle hopping and so account for the observed\cite{jl1} quasi-static intrinsic charge\cite{go1,go3} and spin\cite{ru1,lo1} disorder of cuprates, reflected in the unusual behavior of transport\cite{bo1,ma1,ru1} coefficients and NMR relaxation rates.\cite{go3,jl1}

All this motivates us to investigate carefully the $U_d\gg\Delta_{pd}$ limit of the $t_{pp}$-extended Emery model, omitting at present the electron-phonon coupling (although it is possibly strong\cite{bs2}). In this endeavour, we are led to some extent by the translationally invariant $U_d=\infty$ mean field slave boson (MFSB) theory,\cite{ko1} which predicts for optimal dopings\cite{mr1} the band picture with weak renormalization of the band parameters $t_{pd}$ and $\Delta_{pd}$ ($t_{pp}$ is unaffected\cite{qs1,tu1,mr1}) in the regime $\Delta_{pd}^2\gtrsim2t_{pd}^2\gtrsim \Delta_{pd}|t_{pp}|\gtrsim 4t_{pp}^2$, identified above on neglecting their renormalizations and keeping them fixed for a given parent material. At $x=0$ the $t_{pp}=0$ MFSB describes\cite{ko1} the change of an insulator into a correlated metal through the Brinkman-Rice (BR) phase transition between the $n_d^{MFSB}=1$ and $0<n_d^{MFSB}<1$ states for $(\Delta_{pd}/t_{pd})_{BR}\approx4.7$.\cite{ko1,tu1} This transition is conserved\cite{mr1} for small $t_{pp}$ but shifted linearly in $t_{pp}$ to higher values of $(\Delta_{pd}/t_{pd})_{BR}$. In the MFSB language, the cuprates with $\Delta_{pd}^2\gtrsim2t_{pd}^2\gtrsim\Delta_{pd}|t_{pp}|\gtrsim4t_{pp}^2$ fall below\cite{ku1,sk2} the BR transition at $x=0$ (i.e. $\Delta_{pd}/t_{pd}<(\Delta_{pd}/t_{pd})_{BR}$). For $x$ finite, the MFSB smoothes out the BR transition in $n_d^{MFSB}(x)$. Close below the BR transition a few percent doping of the $x=0$ state can then easily produce\cite{ku1} a sizeable decrease of $n_d^{MFSB}$ from the $x=0$ value $n_d^{MFSB}\approx1$, i.e. $\partial n_d^{MFSB}/\partial x<0$ in the BR regime. Further below the BR transition, the renormalizations become weak and the weakly renormalized metallic $\partial n_d^{MFSB}/\partial x>0$ regime is obtained.\cite{ku1} $\partial n_d^{MFSB}/\partial x=0$ conveniently defines the position $(\Delta_{pd}/t_{pd})_{cs}<(\Delta_{pd}/t_{pd})_{BR}$ of the crossover and gives the corresponding doping value $x_{cs}^{MFSB}(\Delta_{pd}^{cs},t_{pd}^{cs},t_{pp}^{cs})$. Typically,\cite{ku1}   $n_d^{MFSB}(x_{cs}^{MFSB})$ is close to $3/4$ already for $t_{pp}=0$. The $x=0$ value $n_d^{MFSB}$ is, however, somewhat overestimated in the BR regime keeping in mind that the BR transition should itself be smoothed\cite{mr1} in a theory better than MFSB. In particular, the MFSB does not contain the d$^{10}$$\leftrightarrow$d$^9$ disorder and the magnetic correlations. The latter give additional stability\cite{bbk} to the insulating phase and open the possibility of crossing over from the AF-insulating regime to the metallic regime. Moreover, the local gauge invariance, although satisfied on average in the MFSB,\cite{ko1} is irremediably broken.\cite{li1,mr1} Altogether, this leads us to search beyond the MFSB.

Extension of the MFSB is usually attempted from the $x=0$, Mott-AF side, starting from the unperturbed $n_d^{(0)}=1$ N\' eel ground state. That state is widely used to approach the propagation of the first additional hole\cite{za1} or electron\cite{xi1} in the x=0 AF phase of cuprates. The doped hole is placed\cite{za1,ch1} on the {\it upper} oxygen level, assuming that $n_d$, associated with the {\it lower} copper level, is close to unity. Originally, the $t-J$ model with the large $U_d$ superexchange $J_{pd}=4t_{pd}^4/\Delta_{pd}^3$ was so obtained\cite{ge1,za1,tu1,fu1}  for $U_d>\Delta_{pd}\gg t_{pd}>0$ and $t_{pp}=0$. In the opposite limit $U_d>t_{pd}\gg\Delta_{pd}\geq0$ appreciable $t_{pd}$ hybridization within the CuO$_2$ unit cells leads for $x=0$ to a reduction of $n_d$, typically to $n_d\approx1/2$. It is then preferable to put\cite{bb1,sk4}  the additional hole on the {\it lower} (rather than upper) covalent level, let its average charge in the unit cell be shared equally between the Cu and O states, worry about the single particle hybridization among the unit cells, ensure that holes on Cu sites avoid each other {\it in the temporal dimension} and allow for the spin polarization on Cu and O sites, as emphasized in the preliminary report\cite{bb1} of the present work. This means that the $t-J$ model is not a suitable starting point for $t_{pd}\gg\Delta_{pd}$, i.e. that the covalent phase, as described here with dynamic spin and charge disorder, including weak magnetic correlations, prevails then even at $x=0$.
      
The intermediary regime $\Delta^2_{pd}\geq2t_{pd}^2\geq\Delta_{pd}|t_{pp}|\geq4t_{pp}^2$ requires additional care. This regime is apparently at the brink of instability of the coherent $x=0$ Mott-AF phase. The crossover is then expected\cite{bb1} to occur already for small $x\approx x_{cs}$. Nevertheless, the $t-J$ approach is often rigidly extended to all $x>0$ of interest, taking that additional holes go to {\it upper} covalent levels upon tacit assumption that $x<x_{cs}$ in cuprates. Adding phenomenological next-to-next-Cu-neighbor effective hoppings $t'$, $t''$\ldots and, sometimes,\cite{cl1} next-to-next-Cu-neighbor super exchanges $J'$, $J''$ is obviously insufficient in this respect, essentially because  intra- and inter-cell Cu-O covalence\cite{wl1,bb1} of the {\it lower} level and the associated temporal incoherencies\cite{bb1} are omitted.  The restriction to rigid $t$, $t'-J$, $J'$ models can {\it in principle} be relaxed and the $x<x_{cs}$ theory based on the $n_d^{(0)}=1$ N\' eel unperturbed ground state of the $U_d=\infty$ Emery model extended to dopings $x>x_{cs}$ but this requires high order calculations. We are thus tempted to restrict the rigid $t-J$ approach to the range $x<x_{cs}$, including\cite{bbk} the required\cite{wl1} covalent corrections in that limit, while here we describe the doping range $x>x_{cs}$ in terms of a {\it renormalized} three-band theory, determining $x_{cs}$ from the upper side. Such an approach bears some resemblance with nonmagnetic MFSB,\cite{ko1,mr1} NCA,\cite{ni1,tu1} DMFT\cite{zl1,mc1,rb1} and with LDA+U\cite{pe1} calculations for cuprates. It contains dynamic charge and magnetic fluctuations on Cu and O sites, including perturbatively the resonating valence bonding (RVB) related to the generalized superexchange $J_{pd}$.

\section{\texorpdfstring{$U_d=\infty$ perturbation}{Perturbation} theory in terms of \texorpdfstring{$t_{pd}$}{tpd}}

The $T=0$ diagrammatic expansion in terms of $t_{pd}$ is used for this purpose through the slave particle mapping, which is asymptotically locally gauge invariant. The time structure of the perturbation theory plays an essential role because $U_d=\infty$ is replaced by time delays\cite{ka1,fr3,bb1,bbk}, induced in the motion of holes across Cu sites by the propagation of the intermittently added single particle. The corresponding $n_d^{(0)}=1$ slave particle theory is quite intricate, especially\cite{bbk} for $t_{pp}\neq0$, as also indicated by recent small cluster calculations.\cite{sk4} Here we start therefore immediately with the metallic phase for appreciable doping. {\it Partial} infinite sums in terms of $t_{pd}$ are selected with the regime $2t_{pd}^2\gtrsim \Delta_{pd}|t_{pp}|\gtrsim4t_{pp}^2$ in mind. $t_{pd}$ is not the small parameter of those expansions but rather, on noting that the effects of large $U_d$ are reduced if $n_d$ is small, the small parameter is $n_d\leq1/2$.  The effective repulsion, which is proportional to $t_{pd}^4$ (as is the attractive superexchange in the opposite limit), turns out to be small to comparable to $t_{pd}$, provided that $n_d<1/2$. The metallic $n_d\approx1/2$ regime is thus reached in reasonably low order of our diagrammatic Dyson perturbation theory, which gives a practical value to these summations. It is shown from the $x>x_{cs}$ side that $x_{cs}(\Delta_{pd},t_{pd},t_{pp})$ bears then some resemblance to $x_{cs}^{MFSB}(\Delta_{pd},t_{pd},t_{pp})$. We also show how the d$^{10}$$\leftrightarrow$d$^9$ disorder and the SDW correlations enter the single particle propagation, while the singular properties of the coherent O$_x$/O$_y$ and SDW correlations, associated with imperfect nesting, already briefly described elsewhere,\cite{bb1} will be further discussed within the present $x>x_{cs}$ approach in an upcoming publication.\cite{sk3}

The $U_d=\infty$ theory with auxiliary (slave) particles\cite{bn1,co1,ho1} is well known and will be discussed only briefly here. The d$^{10}$ state on Cu at the position $\vec R$ is denoted by $f_{\vec R}^\dagger|\tilde0\rangle$ and the d$^9$ state with spin $\sigma$ by $b^{\sigma\dagger}_{\vec R}|\tilde 0\rangle$, where $|\tilde 0\rangle$ is the auxiliary vacuum on Cu. In the so spanned three-state space (d$_8$ state at $2\varepsilon_d+U_d$ omitted), the number operators of the slave particles satisfy $Q_{\vec R}=n_{f\vec R}+\sum_{\sigma}n^\sigma_{b\vec R}=1$. The physical fermion $c^{\sigma\dagger}_{\vec R}$ projected on the d$^9$, d$^{10}$ subspace is written as $c^{\sigma\dagger}_{\vec R}\rightarrow b^{\sigma\dagger}_{\vec R}f_{\vec R}$. The corresponding number operators satisfy $n^\sigma_{d\vec R}=n^\sigma_{b\vec R}$, usually called the Luttinger sum rule (LSR). $b^{\sigma\dagger}_{\vec R}$ and $f_{\vec R}^\dagger$  can be taken respectively as fermions and bosons ("slave boson theory", SBT) or as bosons and spinless fermions ("slave fermion theory", SFT) in order to satisfy the anticommutation rules on and among Cu sites projected on the d$_9$, d$_{10}$ subspace. The states on oxygens are associated with physical $p$-fermions. The $U_d=\infty$ Hamiltonian written in terms of the auxiliary particles is locally gauge invariant, and commutes with $Q_{\vec R}$, i.e. $Q_{\vec R}=1$ is a physical constant of motion.
 
Therefore we start the time-dependent perturbation theory, in terms of $H_I(t_{pd})$, from the unperturbed, $t_{pd}=0$, $Q_{\vec R}=1$, paramagnetic, translationally invariant slave particle ground state associated with the unperturbed Hamiltonian $H_{0\lambda}-\lambda N=H_{0d}(\varepsilon_d)+\lambda \sum_{\vec R}(Q_{\vec R}-1)+H_{0p}(\varepsilon_p,t_{pp})$

\begin{equation}
|G_{0\lambda}\rangle=|G_0^b\rangle\otimes|G_0^f\rangle\otimes |G_0^p\rangle\label{Eq001}\;,
\end{equation}

\noindent where 
			                                                                              \begin{equation}
|G_0^f\rangle=\prod_{\vec R}f_{\vec R}^\dagger|\tilde 0\rangle\label{Eq002a}
\end{equation}

\noindent is the $n_f^{(0)}=1$ state. The corresponding energy $\lambda$ is the site energy of the $f$-particle appearing, as usual, by adding $\lambda(Q_{\vec R}-1)$ into the unperturbed Hamiltonian $H_0$. In the present perturbation theory $\lambda$ serves as a parameter which checks the local gauge invariance, rather than as the Lagrange multiplier familiar from the MFSB theories.

$|G_0^b\rangle$ is the state with no $b$-particles, i.e. $n_b^{(0)}=n_d^{(0)}=0$. Thus $n_d^{(0)}=0$ is the outset of our expansion in terms of $n_d$ small. Since $|G_0^b\rangle$ is the no-particle state it is nondegenerate, irrespective of the Pauli symmetry of the $b_\sigma$-particles.

The $f$-particles in $n_f^{(0)}=1$ Eqs.~(\ref{Eq001},\ref{Eq002a}) can also be chosen as bosons or spinless fermions. The advantage to choose $f$'s as bosons is that then $b_\sigma$'s can be taken as fermions indistinguishable from $p$-fermions. In the slave particle representation this permits one to satisfy the anticommutation rules between the physical $c$-fermions on the Cu sites and the $p$-fermions on the O-sites. The disadvantage is that the state (\ref{Eq002a}) with bosons is highly degenerate with respect to multiple boson occupations of Cu sites. As is well known, degeneracy of the unperturbed ground state causes problems in time-dependent perturbation theory. This difficulty is obviously eliminated on choosing $f$'s as spinless fermions and $b_\sigma$'s as bosons, since the state of Eq.~(\ref{Eq002a}) is then nondegenerate. However, in that case, we are dealing with three kinds of distinguishable particles, i.e. the anticommutations between the Cu sites and O-sites are replaced by commutations. The corresponding time-dependent perturbation theory (SFT) must be therefore antisymetrized {\it a posteriori} (it will be then named here "modified" SFT, MSFT) and this is the route chosen henceforth.

As the state of Eq.~(\ref{Eq002a}) is nondegenerate for spinless fermions it can be simply expressed in terms of Fourier transforms $f_{\vec k}^\dagger$ of the local operators $f_{\vec R}^\dagger$. Up to an unimportant phase factor we have

\begin{equation}
|G_0^f\rangle=\prod_{\vec R}f_{\vec R}^\dagger|\tilde 0\rangle=
\prod_{\vec k}f_{\vec k}^\dagger|\tilde 0\rangle\;,\label{Eq002b}
\end{equation}

\noindent where the product over $\vec k$ extends over the whole CuO$_2$ Brillouin zone, which corresponds to $N$ CuO$_2$ unit cells. In other words, the Mott state of spinless fermions is equivalent to the full (dispersionless) band of those particles.

Let us finally mention that $|G_0^p\rangle$ is the usual (nondegenerate) Hartree-Fock (HF) state of the $p$-fermions in the cosine band associated with $\varepsilon_p$ and $t_{pp}$. This band contains $2n_p^{(0)}=1+x$ fermions, associated with the chemical potential $\mu_{1+x}$. For $x$ small this band is nearly quarter filled. It is usually folded artificially into the CuO$_2$ Brillouin zone in two $l,\tilde l$ oxygen bands, anticipating the effect of $H_I(t_{pd})$, which is expected to generate three separate bands and make the lowest one nearly half filled. 

Once the unperturbed ground state is expressed\cite{ni1,mr1} in the full momentum representation so should the slave particle Hamiltonian $H_\lambda=H_{0\lambda}-\lambda N+H_I$. We have, in terms of $f_{\vec k}^\dagger$, $b^\dagger_{\vec k,\sigma}$, $p^{(i)\dagger}_{\vec k,\sigma}$ ($i=l,\tilde l$) and their hermitean conjugates, 

\[H_{0\lambda}=\sum_{i,\vec k,\sigma}\varepsilon_{p\vec k}^{(i)}p^{(i)\dagger}_{\vec k,\sigma} p^{(i)}_{\vec k,\sigma}
+\sum_{\vec k,\sigma}(\varepsilon_d+\lambda)b^\dagger_{\vec k,\sigma} b _{\vec k,\sigma}+\lambda\sum_{\vec k}f^\dagger_{\vec k}f_{\vec k}\]

\begin{eqnarray}
H_I&=&\frac{i}{\sqrt N}\sum_{i,\sigma,\vec k,\vec q}
t^{(i)}_{pd}(\vec k)b_{\vec k+\vec q,\sigma}^\dagger f_{\vec q}
p_{\vec k,\sigma}^{(i)}+\hc\label{Eq003} \\
t_{pd}^{(i)}(\vec k)&=&t_{pd}\sqrt 2
\left(|\sin{\frac{k_x}{2}}|\pm|\sin{\frac{k_y}{2}}|\right)\;.\nonumber
\\\nonumber 
\end{eqnarray}

\noindent assuming the D$_4$ symmetry. Here, $t_{pd}^{(i)}(\vec k)$ describes the fact that by annihilating the $f_{\vec q}$ spinless fermion and by creating the $b^\dagger_{\vec k+\vec q,\sigma}$ boson, one annihilates the $p_{\vec k,\sigma}^{(i)}$ fermion in either of two $i=l,\tilde l$ bands $\varepsilon_{\vec k}^{(i)}=\varepsilon_p\pm4|t_{pp}|\sin{(k_x/2)}\sin{(k_y/2)}$. $\lambda$ is introduced in Eq.~(\ref{Eq003}) as the test parameter (rather than as the Lagrange multiplier) since the physical $Q_{\vec R}=1$ result, independent of $\lambda$, must be ultimately achieved. As will be seen below the present time-dependent perturbation theory will prove independent of $\lambda$ in each order.
 
The perturbation theory can then be carried out in terms of $H_I$ on the top of the nondegenerate state of Eqs.~(\ref{Eq001},\ref{Eq002a},\ref{Eq002b}), the time orderings and the normal orderings being well defined together with the Pauli symmetry of the relevant $b$-, $f$-, and $p$-particles. Since both the Hamiltonian $H_\lambda$ and the unperturbed ground state are locally gauge invariant, translationally invariant on the CuO$_2$ lattice, and symmetric under time reversal, the SFT will either generate the exact ground state with the same symmetries, or break them in a controlled way. It is thus left to the SFT to keep $Q_{\vec R}=1$, generate the LSR $n_d=n_b$, obey the anticommutation rules on and among the Cu sites, and to satisfy the charge conservation rule $n_d+2n_p =1+x$. Such a multiband SFT does not suffer from problems related to the breakdown of local gauge invariance, encountered\cite{li1} in the single band models. The MSFT is eventually constructed only to take care of the Cu-O anticommutation rules, as well as possible.

We wish to emphasize the intimate relation between local gauge invariance and causality in the time-dependent $T=0$ perturbation theory. The latter has therefore advantages over the finite-$T$ Matsubara theory often encountered\cite{ni1,fu1} in the slave particle context. The Matsubara theory uses the canonical ensemble in the full slave particle space and therefore treats the $Q_{\vec R}=1$ and the $Q_{\vec R}\neq1$ states on equal footing, provided that they are degenerate in energy. It is therefore more difficult to control the $Q_{\vec R}=1$ local gauge invariance in the finite $T$ Matsubara theory than in the time-dependent $T=0$ approach.

The elementary bricks which build the time-dependent perturbation theory according to Wick's theorem are the free-particle propagators. Defining, as usual, $B_\lambda(\vec k,t) = 
-i\langle T b_{\vec k}b_{\vec k}^\dagger(t)\rangle$, we find that the free propagator of the $b$-particle is dispersionless, 
   
\begin{equation}
B_\lambda^{(0)}=\frac{1}{\omega-\varepsilon_d-\lambda+i\eta}\;.\label{Eq004}
\end{equation}

\noindent Through $+i\eta$ it describes the intermittent creation of the $b$-particle, while its annihilation is impossible in the no-bosons state of Eq.~(\ref{Eq001}). In contrast, the spinless fermions can only be annihilated, i.e. $F_\lambda^{(0)}$ is obtained from Eq.~(\ref{Eq004}) by replacing $\varepsilon_d+\lambda$ by $\lambda$ and $+i\eta$ by $-i\eta$. The free propagators of $1+x$ $p$-particles contain both $+i\eta$ and $-i\eta$ components $G_p^{(i)>}(\vec k,\omega)$ and $G_p^{(i)<}(\vec k,\omega)$ according to their Fermi distribution $f_{\vec k}^{(i)}$ in the HF state associated with the $i=l,\tilde l$ bands. The corresponding chemical potential is hereafter denoted by $\mu^{(0)}$. With $x<2$, only the states in the $l$-band are occupied.

\begin{figure}[htb]

\begin{center}{\scalebox{0.5}
{\includegraphics{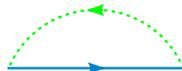}}}
\end{center}

\caption{(Color online) Propagator $\LD^{(0)}$ of the spinless fermion (green)-boson (blue)  pair carrying wave vector $\vec k$ and energy $\omega$; arrows denote that the $b$-particle (blue) can only advance and the $f$-particle (green) recede in time.\label{fig001}}

\end{figure}

The $d$-particle propagator is defined by $\LD_{\vec k}(t)=-(i/N)\langle T\sum_q f_q^\dagger b_{k+q}f_q(t) b_{k+q}^\dagger(t)\rangle$. $\LD_{\vec k}^{(0)}$, shown diagramatically in Fig.~\ref{fig001}, is thus also dispersionless and obtained from Eq.~(\ref{Eq004}) by replacing $\varepsilon_d+\lambda$ by $\varepsilon_d$. $\LD_{\vec k}^{(0)}=(\omega-\varepsilon_d+i\eta)^{-1}$ thus firstly reproduces $n_d^{(0)}=0$ and secondly is independent of $\lambda$, as any physical quantity should be. In the next step we define $\LD_{\vec k}^{(r)}$ associated with the $r$-th order time-dependent perturbation theory $r>0$. According to the definition of a particle-hole pair, $\LD_{\vec k}^{(r)}(\omega)$ is given by the (generalized)  Bethe-Salpeter equation

\begin{equation}
\LD_{\vec k}^{(r)}(\omega)=\GD_{\vec k}^{(r-1)}(\omega)+\GD_{\vec k}^{(r-1)}(\omega)\Gamma_{\vec k}^{(r)}(\omega)\GD_{\vec k}^{(r-1)}(\omega)\;.\label{Eq005a}
\end{equation}

\begin{figure}[htb]

\begin{center}{\scalebox{0.5}
{\includegraphics{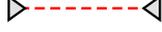}}}
\end{center}

\caption{(Color online) Four-leg vertex $\Gamma_{\vec k}^{(0)}(\omega)$; triangular vertices are $t_{pd}^{(i)}(\vec k)$ and the red lines are the free propagators $G_p^{(i)}(\vec k,\omega)$ combined according to Eq.~(\ref{Eq006}).  \label{fig002}}

\end{figure}

\noindent Here $\GD_{\vec k}^{(r-1)}$ is the quantity irreducible with respect to cutting the $p$-lines and $\Gamma_{\vec k}^{(r)}(\omega)$ is the renormalized  four-leg vertex given iteratively by the Dyson equation

\begin{equation}
\Gamma_{\vec k}^{(r)}(\omega)=\Gamma_{\vec k}^{(0)}(\omega)+\Gamma_{\vec k}^{(0)}(\omega)\GD_{\vec k}^{(r-1)}(\omega)\Gamma_{\vec k}^{(r)}(\omega)\;.\label{Eq005b}
\end{equation}

\noindent in terms of the bare four-leg vertex $\Gamma_{\vec k}^{(0)}(\omega)$ characterized by $\mu^{(0)}$, shown in Fig.~\ref{fig002},

\begin{equation}
\Gamma_{\vec k}^{(0)}(\omega)=[t_{pd}^{(l)}(\vec k)]^2G_p^{(l)<}(\vec k,\omega)+\sum_{i=l,\tilde l}[t_{pd}^{(i)}(\vec k)]^2G_p^{(i)>}(\vec k,\omega)\;.\label{Eq006}
\end{equation}

\noindent Eq.~(\ref{Eq006}) generalizes the four-leg vertex used previously\cite{ni1} for $t_{pp}=0$.

Actually, $t_{pd}^{-2}\Gamma_{\vec k}^{(r)}$ in Eq.~(\ref{Eq005b}) can be interpreted as an appropriately symmetrized generalization to the Emery model of the wide-band propagator on the Anderson lattice. According to Eq.~(\ref{Eq006}) the intermittent $p$-particle $t_{pd}^{-2}\Gamma_{\vec k}^{(0)}$ is prepared in {\it two} $i=l,\tilde l$ $p$-bands, instead of one. $t_{pd}^{-2}\Gamma_{\vec k}^{(r)}$ is thus the canonical $pdp$ propagator which appears naturally in the perturbation theory for the Emery model. For $r>0$, this propagation involves both $i=l,\tilde l$ bands and the d-state, similarly to Eq.~(\ref{Eq005a}) for an intermittently created $d$-particle. Eq.~(\ref{Eq005b}) can thus be interpreted as the Dyson equation for the $t_{pd}^{-2}\Gamma_{\vec k}^{(r)}$ single particle $pdp$ propagator with the Dyson self energy $\Pi^{(r)}=t_{pd}^2\GD_{\vec k}^{(r-1)}$ irreducible with respect to $t_{pd}^{-2}\Gamma_{\vec k}^{(0)}$-lines. According to Eq.~(\ref{Eq003}) for $H_I$, the lowest order $\GD_{\vec k}^{(0)}$ is simply equal to $\LD^{(0)}$ of Fig.~\ref{fig001}. In other words $\LD^{(0)}$ is not only the elementary $d$-particle propagator but also the essential component of the lowest order "local" irreducible self energy $\Pi^{(1)}=t_{pd}^2\LD^{(0)}$ in Eq.~(\ref{Eq005b}) for $t_{pd}^{-2}\Gamma_{\vec k}^{(1)}$. The $r=1$ procedure thus separates out the $\vec k$-independent free $d$-propagator $\LD^{(0)}$ in the leading $pdp$-particle self energy $\Pi^{(1)}$ on associating in Eq.~(\ref{Eq006}) the $\vec k$-dependence of triangular vertices of Fig.~\ref{fig002} with the $\vec k$-dependent free $p$-propagator $t_{pd}^{-2}\Gamma_{\vec k}^{(0)}$.

On the other hand $\LD_{\vec k}^{(r)}(\omega)$ describes the creation/annihilation of the intermittent $d$-particle on the Cu sites and its subsequent $dpd$ propagation. The factors $\GD_{\vec k}^{(r-1)}(\omega)$ in Eq.~(\ref{Eq005a}) are the same as those involved in $\Gamma_{\vec k}^{(r)}$. Due to this $\GD_{\vec k}^{(r-1)}$ can be taken as the effective "free" propagator on the Cu-site, which allows one to perform the resummation of the Bethe-Salpeter Eq.~(\ref{Eq005a}) into the Dyson form, in accordance with the idea that $\LD_{\vec k}^{(r)}$ is the propagator of the single $dpd$ particle. Nevertheless, $\GD_{\vec k}^{(r-1)}$ is not {\it in general} a fermion propagator and, consequently, neither is $\LD_{\vec k}^{(r)}$.

The reason is that the SFT removes the d$^8$ state completely, while the original large $U_d$ theory treats it as the {\it empty} state (upper Hubbard band) at large energy $\varepsilon_d+U_d$. Thus, in order to satisfy the equality $n_{d+}^{(r)}=n_{d-}^{(r)}$ associated with the full fermion anticommutation rule on the Cu site and that obtained from $\LD_{\vec R}^{(r)}(t)$ in the $t\rightarrow0_\pm$ limits, one should allow for the additional spectral density at the energy $\varepsilon_d+U_d$ on the $t<0$ side. This adds a term

\begin{equation}
\sigma^{(r-1)>}=\frac{a^{(r-1)}}{\omega-(\varepsilon_d+U_d)+i\eta}\;,\label{Eq007}
\end{equation}

\noindent to $\GD^{(r-1)}$, which takes explicitly into account the fact that the d$^8$ state is empty. Although permanent (average) occupation of the d$^8$ state is forbidden in the $U_d\rightarrow\infty$ limit, the full fermion nature of $d$-particles requires visits of the d$^8$ state on the $t<0$ side. The point is that when the expression (\ref{Eq007}) is integrated over $\omega$ in the Fourier transform which determines the overall spectral density associated with the $t<0$ component of $\GD^{(r-1)}$ it gives a contribution $a^{(r-1)}$ which has to be retained in spite of $U_d\rightarrow\infty$. In principle, $a^{(r-1)}$ may be determined from the anticommutation requirement $n_{d+}^{(r)}=n_{d-}^{(r)}$ of the extended Eq.~(\ref{Eq005a}) and is expected to be small for $n_{d+}^{(r)}$ small. In practice, following the spirit of the slave particle theories, we associate the physical average occupation $n_d^{(r)}$ with $n_{d+}^{(r)}$ of Eq.~(\ref{Eq005a}), not worrying about $\sigma^{(r-1)>}$ at all.

The last step is to determine the chemical potential $\mu^{(r)}$ of the $p$-fermions in order to satisfy the conservation of the total charge $1+x$. In the SFT this can be done by requiring $n_b^{(r)}+2n_p^{(r)}=1+x$, with the number of bosons $n_b^{(r)}$ found below. Alternatively, one can require $n_d^{(r)}+2n_p^{(r)}=1+x$ on keeping in mind that the LSR $n_b=n_d$, though approximate in low orders, holds asymptotically in the exact SFT.

\section{Leading correction to single particle \texorpdfstring{$p$}{p}- and \texorpdfstring{$d$}{d}-propagators\label{Sec04}}

Here we choose the $n_d^{(r)}+2n_p^{(r)}=1+x$ prescription because, as we shall see now, Eqs.~(\ref{Eq005a}-\ref{Eq006}) at $r =1$ then become equivalent to the HF equations for the free propagators in the $tpd$ hybridized $U_d=0$ model. The deep reason for this far-reaching result is that the Cu site is initially empty, Eqs.~(\ref{Eq001}-\ref{Eq002b}), and at the order $r =1$ the intermittent particle does not probe two-particle effects which involve the Cu site.

More formally, $\LD_{\vec k}^{i}$ involves $\GD_{\vec k}^{(0)}=\LD_{\vec k}^{(0)}=(\omega-\varepsilon_d+i\eta)^{-1}$, which is associated with unspecified commutation properties through the absence of the $-i\eta$ component. The single particle problem of anticrossing\cite{mr1} between the $\varepsilon_d$ level and the "two" oxygen $i=l,\tilde l$ bands is then solved exactly by $\LD_{\vec k}^{(1)}$ and $\Gamma_{\vec k}^{(1)}$ of Eqs.~(\ref{Eq005a}-\ref{Eq006}). Both $t_{pd}^{-2}\Gamma_{\vec k}^{(1)}$ and $\LD_{\vec k}^{(1)}$ in the Dyson form exhibit coherent poles belonging to three bands (branches of poles) $\omega_{\vec k}^{(j)}$ denoted respectively by $j=L,I,U$. The poles $\omega_{\vec k}^{(j)}$ in the HF $t_{pd}^{-2}\Gamma_{\vec k}^{(1)}$ are associated with the residuals (spectral weights) $z_{\vec k}^{(j)}(\omega_{\vec k}^{(j)})$ which can be expressed entirely in terms of the three $\omega_{\vec k}^{(j)}$. For example, for the lowest band $L$

\begin{equation}
z_{\vec k}^{(L)}(\omega_{\vec k}^{(L)})=\frac{(\omega_{\vec k}^{(L)}-\varepsilon_d)^2(\omega_{\vec k}^{(L)}-\varepsilon_{p\vec k}^{(l)})(\omega_{\vec k}^{(L)}-\varepsilon_{p\vec k}^{(\tilde l)})}{t^2_{pd}(\omega_{\vec k}^{(I)}-\omega_{\vec k}^{(L)})(\omega_{\vec k}^{(U)}-\omega_{\vec k}^{(L)})}
\label{Eq008}\;,
\end{equation}

\noindent and similarly for other two bands. The chemical potential $\mu^{(1)}$ of the $p$-fermions is next defined as the energy which separates the poles of the $p$-propagator $t_{pd}^{-2}\Gamma_{\vec k}^{(1)}$ in the upper and lower $\omega$-plane. Since the same poles appear in $\LD_{\vec k}^{(1)}$, this step accounts for them too, with no reference to the Pauli symmetry of the $b^\dagger f$ pairs.  In contrast to $\mu^{(0)}$, which defines the average number of $p$-fermions on the O-sites and allows for their fluctuations among those sites with the total number of $p$-fermions fixed, $\mu^{(1)}$ allows also for fluctuations of the total number of $p$-fermions by their conversion into the $b^\dagger f$ pairs. $\mu^{(1)}(\Delta_{pd}, t_{pd}, t_{pp}, x)$ is determined through the approximate charge conservation rule $n_d^{(1)}+2n_p^{(1)}=1+x$, bearing in mind that $n_p^{(1)}$ and $n_d^{(1)}$ are defined by Eqs.~(\ref{Eq005a},\ref{Eq005b}) as functions of the band parameters and $\mu^{(1)}$. The whole $r =1$ procedure described above thus amounts to the redistribution of the spectral weights and the Fermi occupation factors $f_{\vec k}^{(i)}$ (with accompanying $\pm i\eta$'s) from two oxygen bands $i=l,\tilde l$ and the empty $d$-state into the three hybridized bands $\omega_{\vec k}^{(j)}$, $j=L,I,U$. In other words, the unperturbed ground state of Eq.~(\ref{Eq001}) evolves through the prescription $n_d^{(1)}+2n_p^{(1)}=1+x$ into the HF state of the coherently hybridized {\it noninteracting} ($U_d=0$) $pd$ particles with a shift $\mu^{(1)}-\mu^{(0)}$ in the chemical potential from the upper to the lower hybridized hole states.  The shift is {\it large} when single particle anticrossing is important. However it is immediately evident that local gauge invariance is not obeyed for $r =1$, because the double occupation of the Cu-site is allowed in the $pd$ hybridized HF state. This will be corrected in higher orders. 

Concerning the anticommutation rule on the Cu site, we can take the $t\rightarrow0$ limit of $\LD_{\vec R}^{(1)}(t)$ in the Dyson form, to find that $n_{d-}^{(1)}=n_{d+}^{(1)}=n_d^{(1)}$, i.e. $a^{(0)}=0$ in Eq.~(\ref{Eq007}). The anticommutation rule on the Cu site is thus satisfied. However it is immediately evident that local gauge invariance is not obeyed for $r=1$, because the double occupation of the Cu-site is allowed in the $pd$ hybridized HF state. Later we shall return to this point more formally.

Although the above argument, which shows that the $U_d=\infty$ SFT generates the $U_d=0$ HF result in the lowest order, does not require explicitly that $n_d^{(1)}$ is small, the SFT, which starts with $n_d^{(0)}=0$,  will converge quickly to satisfy local gauge invariance only when this condition is met. Let us therefore mention briefly the values of the single particle parameters $\Delta_{pd}$, $t_{pd}$, and $t_{pp}$ which make $n_d^{(1)}(\Delta_{pd}, t_{pd}, t_{pp},x)$ small for a given $1+x$. These conditions can be taken over directly from the three band HF theory\cite{mr1} which determines $n_d^{(1)}$ from the partial derivative of the HF energy of Eq.~(\ref{Eq003}) with respect to $\varepsilon_d$, thus circumventing the clumsy calculation via the spectral density of $\LD^{(1)}$ complementary to that of Eq.~(\ref{Eq008}).

The simplest situation\cite{fr1,fr3,bs5} corresponds to $t_{pd}\gg\Delta_{pd}$ and $t_{pp}=0$, where one immediately finds $n_d^{(1)}=1/2$ taking formally $x=0$ (having the metallic regime with small $x>x_{cs}$ in mind). Note, in this respect, that $\mu^{(1)}$ coincides for $x=0$ with the vH energy $\omega_{vH}$ in the lowest $L$-band and that the latter is almost independent of $\varepsilon_d$ in the limit considered. This results in the equal sharing of the charge between one Cu and {\it two} O's which, indeed, was traditionally obtained\cite{fr1,fr3,bs5} in this way.

Finite $t_{pp}$ can be easily included in this scheme perturbatively for $|t_{pp}|\ll t_{pd}$, $\Delta_{pd}$. In contrast to $t_{pd}$, $t_{pp}$ shifts the Fermi level $\mu^{(1)}$ at $x=0$ from the vH singularity at $\omega_{vH}$.  A finite doping $x=x_{vH}$ is therefore required to reach the vH singularity. $x_{vH}$ was found\cite{mr1} to be equal ($t_{pp}<0$) to $-32t_{pp}/\pi^2\Delta_{pd}$ in the limit $\Delta_{pd}>t_{pd}$ and the corresponding HF energy was determined analytically. This can be readily extended to the $t_{pd}>\Delta_{pd}$ limit with $-t_{pp}\Delta_{pd}\ll t_{pd}^2$ when $x_{vH}\approx-32t_{pp}\Delta_{pd}/\pi^2t_{pd}^2$. The large numerical factor 32, multiplying $t_{pp}$ in this equation, is due to four $t_{pp}$ bonds per two $t_{pd}$ bonds in the CuO$_2$ unit cell. This compensates, to some extent, the chemical inequality $t_{pd}>|t_{pp}|$. The analytic calculation\cite{mr1} of the HF energy follows the same lines. A small $|t_{pp}|$ is thus found to reduce the value of $n_d^{(1)}$. In particular, for $t_{pd}\gg\Delta_{pd}$ the $x=0$ value of $n_d^{(1)}$ falls below $1/2$.

These results can be further extended to the physical regime\cite{mr1} with sizeable $t_{pp}$, satisfying $2t_{pd}^2\approx-t_{pp}\Delta_{pd}$. This regime also sets $\mu^{(1)}$ in the $L$-band, while, importantly, the $I$-band is flat at $\varepsilon_p$. $t_{pp}$ is quite efficient in reducing $n_d^{(1)}$ below $1/2$ at small $x$, due to a large numerical factor carried by $t_{pp}$ in this regime as well. Although the roles of $t_{pd}$ and $t_{pp}$ are comparable in the dispersion, a sizeable $t_{pp}$ requires, in contrast to $t_{pd}$, a sizeable doping $x=x_{vH}$  to reach the vH singularity. In other words $\mu^{(1)}$ for $x=0$ falls well below $\omega_{vH}$ (see e.g. Fig.~\ref{fig004}a), i.e. $n_d^{(1)}$ is reduced appreciably with respect to $1/2$.

The useful overall rule of thumb is that $n_d^{(1)}<1/2$ as long as $\mu^{(1)}$ falls below the vH singularity at $\omega_{vH}$ (independent of $t_{pp}$), i.e. as long as $x<x_{vH}$. In other words, the condition $\mu_{cs}^{(1)}\approx\omega_{vH}$ can be used to define the lowest approximation $x_{cs}^{(1)}(\Delta_{pd},t_{pd},t_{pp})$ to the crossover function $x_{cs}(\Delta_{pd},t_{pd},t_{pp})$ although $\partial n_d^{(1)}/\partial x>0$ all over the parameter space of interest.

\section{Approximate local gauge invariance and the LSR}

\begin{figure}[htb]

\begin{center}{\scalebox{0.5}
{\includegraphics{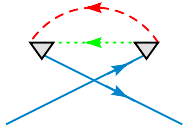}}}
\end{center}

\caption{(Color online) Lowest order renormalization for $B$. The arrows of time are shown. The energy and momentum enter the bubble from right to left.\label{fig003a}}

\end{figure}

\begin{figure}[htb]

\begin{center}{\scalebox{0.5}
{\includegraphics{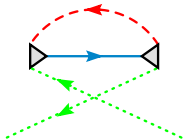}}}
\end{center}

\caption{(Color online) $b\leftrightarrow f$ symmetric lowest order renormalization for $F$.\label{fig003b}}

\end{figure}

Here we turn first to the properties of the propagators $B_{\lambda}^{(1)}(\vec k,\omega)$ and $F_{\lambda}^{(1)}(\vec k,\omega)$, required to asses the accuracy of the above HF result and to construct the next order $r=2$ iteration of Eqs.~(\ref{Eq005a},\ref{Eq005b},\ref{Eq007}). The lowest order connected diagrams for $\Delta B_{\lambda}^{(1)}(\vec k,\omega)$ and $\Delta F_{\lambda}^{(1)}(\vec k,\omega)$ are shown in Figs.~\ref{fig003a},\ref{fig003b}. As usual, the disconnected diagrams, which describe here the incoherent d$^{10}\leftrightarrow d^{9}$ vacuum fluctuations unaffected by the intermittent particles, are not shown. The arrows of time, associated with $\pm i\eta$ factors in elementary propagators $B_{\lambda}^{(0)}(\vec k,\omega)$ and $F_{\lambda}^{(0)}(\vec k,\omega)$ of Eq.~(\ref{Eq004}) (independent of $\vec k$) are depicted in order to emphasize that it is important to account for the full temporal structure of the theory. The bubbles which appear in Figs.~\ref{fig003a} and \ref{fig003b} are the lowest order irreducible Dyson self energies for $B_{\lambda}^{(1)}(\vec k,\omega)$ and $F_{\lambda}^{(1)}(\vec k,\omega)$. They both contain $\Gamma_{\vec k}^{(0)}(\omega)$ of Eq.~(\ref{Eq006}) and involve summation over the occupied states in the $l$-band as indicated in Figs.~\ref{fig003a},\ref{fig003b} by the $p$-propagator going (only) backwards in time. In Fig.~\ref{fig003a} the external frequency enters the bubble from right to left, while in Fig.~\ref{fig003b} it goes from left to right. In addition, the spin factor 2 multiplies the $f$-bubble, unlike the $b$-bubble, but we remember that there are two $b$-bosons for each $\vec k$. This symmetry, exemplified by Figs.~\ref{fig003a},\ref{fig003b}, will be referred to here as the $f\leftrightarrow b$ symmetry. 

It means that the two diagrams of Figs.~\ref{fig003a},\ref{fig003b}, taken together keep the average value of $Q_{\vec R}$ close to unity. However, as easily seen, $\Delta B_{\lambda}^{(1)}(\vec k,\omega)$ and $\Delta F_{\lambda}^{(1)}(\vec k,\omega)$ are singular when the anticrossing\cite{mr1} between the $\varepsilon_d$ level and bands $i=l,\tilde l$ occurs for the states occupied in the $l$-band. This leads to a large difference between $n_d^{(1)}$ of Eq.~(\ref{Eq005a}), which takes the single particle anticrossing into account exactly, and the number $n_b^{(1)}$ of $b$-bosons, which corresponds to the Schr\" odinger perturbation theory for the occupied single-particle states.  The resulting convergence of the SFT is poor. 

It is therefore essential to re-sum the perturbation theory by associating the receding "oxygen" line in Figs.~\ref{fig003a},\ref{fig003b} with hybridized $\Gamma_{\vec k}^{(1)}$ rather than with $\Gamma_{\vec k}^{(0)}$. For example, the $b$-bubble is $\vec k$-independent ("local"), and given by 

\begin{equation}
\beta_\lambda^{(1)}(\omega)=\frac{t_{pd}^2}{N}\sum_{\vec k}\frac{z_{\vec k}^{(L)}f_{\vec k}^{(L)}}{-\omega+\omega_{\vec k}^{(L)}+\lambda+2i\eta}\label{Eq009}\;,
\end{equation} 

\noindent where $z_{\vec k}^{(L)}$ are the residues of the propagators $t_{pd}^{-2}\Gamma_{\vec k}^{(1)}$ given by Eq.~(\ref{Eq008}). It is also assumed for simplicity that $t_{pd}$ is large enough to have\cite{mr1} only the states in the lowest $L$-band occupied. Eq.~(\ref{Eq009}) is obtained by taking into account the Pauli nature of the involved particles, the energy conservation in each triangular vertex and by noting that both propagators in the $b$-bubble are running backwards in time. The set of occupied poles in Eq.~(\ref{Eq009}) lies in the upper $\omega$-half-plane, which makes $\Delta n_b^{(1)}$, associated with $\Delta B_\lambda^{(1)}$ of Fig.~\ref{fig003a}, finite for $1+x\neq0$, independently of $\lambda$.

We are now in a position to discuss the LSR $n_b=n_d$ to the first order. Although $n_d^{(1)}$ is given by the HF theory and $n_b^{(1)}$ is evaluated in the Appendix it is more instructive to carry their comparison as follows. $\Delta B_{\lambda}^{(1)}$ given by Fig.~\ref{fig003a} with $pdp$ hybridized $\Gamma_{\vec k}^{(1)}$ on the "oxygen" line gives $\Delta n_b^{(1)}$. $n_{d+}^{(1)}$ can be conveniently determined from the second term $\GD^{(0)}\Gamma_{\vec k}^{(1)}\GD^{(0)}$ of the Bethe-Salpeter Eq.~(\ref{Eq005a}) before its resummation into the Dyson form. On noting further that the product of the squared pole and a single pole of such $\GD_{\vec k}^{(1)}$ occurs also in $\Delta B^{(1)}_\lambda$, one finds easily that  
			                 
\begin{equation}
\Delta n_b^{(1)}=n_d^{(1)}\label{Eq010a}\;.
\end{equation}

\noindent Although Eq.~(\ref{Eq010a}) departs from the LSR $n_b=n_d$, the difference between $\Delta n_b^{(1)}$ and $n_b^{(1)}$ is apparently small if $n_d^{(1)}$ is small, as shown in detail in the Appendix. The higher order corrections to $n_b^{(1)}\approx n_d^{(1)}$ can then restore the LSR quickly, as the SFT converges from $Q_{\vec R}^{(1)}\approx1$ towards the local gauge invariance $Q_{\vec R}=1$.

Indeed, a conclusion similar to $n_b^{(1)}\approx n_d^{(1)}$ holds for the average value of $Q_{\vec R}$. A result analogous to $B_\lambda^{(1)}(\omega)$ is obtained for $F_\lambda^{(1)}(\omega)$ by $f\leftrightarrow b$ symmetry. $\phi_\lambda^{(1)}(\omega)$ has a similar structure as complex conjugate of $-\beta_\lambda^{(1)}(\omega)$ of Eq.~(\ref{Eq009}) except that a spin factor $2$ multiplies the sum over the occupied states in the $L$-band.  The exact equation, which is analogous to Eq.~(\ref{Eq010a}), 

\begin{equation}
\Delta n_f^{(1)}=-\Delta n_b^{(1)}\label{Eq010b}\;,
\end{equation}

\noindent is therefore transformed into the approximate relation $Q_{\vec R}^{(1)}\approx1$ for $n_b^{(1)}\approx\Delta n_b^{(1)}=n_d^{(1)}$ small, as further discussed in Appendix. $Q_{\vec R}^{(1)}\approx1$ is then subject to higher order corrections which asymptotically enforce the local gauge invariance $Q_{\vec R}=1$. The discussion of Eqs.~(\ref{Eq010a},\ref{Eq010b}) uncovers thus the mechanism of achieving the local gauge invariance in the SFT. The point emphasized here is that physically significant results are obtained already in the low order SFT provided that the average Cu charge $n_d$ (but not necessarily the total charge $1+x$) is small.

\section{Band narrowing and \texorpdfstring{$\grgd^{10}$$\leftrightarrow$$\grgd^9$}{d10-d9} charge-transfer disorder}

Once $B_\lambda^{(1)}(\omega)$ and $F_\lambda^{(1)}(\omega)$ have been determined, they can be used to calculate $\GD^{(1)}\sim(B_\lambda^{(1)}\ast F_\lambda^{(1)})$ in Eqs.~(\ref{Eq005a}) and (\ref{Eq005b}), i.e. to advance the iteration one step further to find $\LD_{\vec k}^{(2)}$ and $\Gamma_{\vec k}^{(2)}$. Using the relation $B^{(1)}_\lambda(\omega)=B^{(1)}_0(\omega-\lambda)$ which follows from Eqs.~(\ref{Eq004}) and (\ref{Eq009}), and $b\leftrightarrow f$ symmetrically $F^{(1)}_\lambda(\omega)=F^{(1)}_0(\omega-\lambda)$, we find (after integration over $\omega-\lambda$) that $\GD^{(1)}$ is independent of $\lambda$. It is further shown in the Appendix that, according to Eq.~(\ref{Eq009}), $B_\lambda^{(1)}(\omega)$ can be written in terms of one pole in the negative $\omega$-half-plane and a set of poles in the positive $\omega$-half-plane, $B_\lambda^{(1)}=B_\lambda^{(1)>}+B_\lambda^{(1)<}$, and $f\leftrightarrow
b$ symmetrically for $F_\lambda^{(1)}=F_\lambda^{(1)<}+F_\lambda^{(1)>}$ (the superscripts $<$ and $>$ denote arrows of time). As usual (see Eq.~(\ref{A2})) the relevant contributions to the convolution $\GD^{(1)}\sim(B_\lambda^{(1)}\ast F_\lambda^{(1)})$ come from the poles on the opposite sides of the $\omega$-axis,

\begin{eqnarray}
\GD^{(1)}&=&\frac{-i}{2\pi}(B_\lambda^{(1)}\ast F_\lambda^{(1)})\nonumber\\&=&
\frac{-i}{2\pi}(B_\lambda^{(1)>}\ast F_\lambda^{(1)<}+
B_\lambda^{(1)<}\ast F_\lambda^{(1)>})\nonumber\\
&=&\GD^{(1)>}+\GD^{(1)<}\label{Eq011}\;.
\end{eqnarray}

\noindent $\GD^{(1)}$ is "local" (dispersionless) in the direct space, i.e. it "occurs" on the Cu site. As shown in the Appendix, we obtain in this way (independently of $\lambda$)

\begin{eqnarray}
\GD^{(1)>}&=&\frac{(1+n_b^{(1)}/2)n_f^{(1)}}{\omega-
\tilde\varepsilon_d+2i\eta}\label{Eq012a}\\
\GD^{(1)<}&=&\frac{t_{pd}^4}{N^2}\nonumber\\&\times&\sum_{\vec k',\vec k''}f_{\vec k'}^{(L)}f_{\vec k''}^{(L)}\frac{A_{\vec k',\vec k''}(\omega)}{\omega-\tilde\omega_{b\vec k'}^{(L)}
-\tilde\omega_{f\vec k''}^{(L)}+\varepsilon_d-4i\eta}\;.\label{Eq012b}
\end{eqnarray}

\noindent $n_b^{(1)}$, $n_f^{(1)}$, $\tilde\varepsilon_d$, $A_{\vec k',\vec k''}$, $\tilde\omega_{b\vec k'}^{(L)}$, $\tilde\omega_{f\vec k''}^{(L)}$ in Eqs.~(\ref{Eq012a}) and (\ref{Eq012b}) can all be expressed in terms of the spectral weight $z_{\vec k}^{(L)}$ of Eq.~(\ref{Eq008}) and the chemical potential $\mu^{(1)}$. E.g., $\varepsilon_d^{(1)}= \varepsilon_d+3\beta_\lambda^{(1)}(\omega=\varepsilon_d+\lambda)$ where $\beta^{(1)}_\lambda(\omega=\varepsilon_d+\lambda)$ is negative according to Eq.~(\ref{Eq009}). Importantly, $\GD^{(1)}=\GD^{(0)}=\LD^{(0)}$ for $1+x=0$, i.e. the propagation of the intermittent particle in the empty band is free, subject only to the single particle hybridization. $t_{pd}^4$ in Eq.~(\ref{Eq012b}) is exhibited in order to stress that the leading term in the $t_{pd}$ expansion of $\GD^{(1)<}$ is quartic in $t_{pd}$ , in contrast to that of $\GD^{(1)>}$, which is then quadratic in $t_{pd}$ because $n_b^{(1)}$ and $n_f^{(1)}$ are. This clearly demonstrates that $\GD^{(1)}$ of Eq.~(\ref{Eq011}) is not a fermion propagator. $n_b^{(1)}$, $n_f^{(1)}$ as functions of $\Delta_{pd}$, $t_{pd}$, $t_{pp}$ and $x$, as well as the double sum in Eq.~(\ref{Eq012b}), are evaluated in the Appendix in the $N\rightarrow\infty$ limit, where the sets of dense poles are treated as cuts in the $\omega$-plane, once the temporal decomposition of $B_\lambda^{(1)}$, $F_\lambda^{(1)}$ and $\GD^{(1)}$ is properly determined. In particular, the shifts of dense poles can be neglected in the $N\rightarrow\infty$ limit, $\tilde\omega_{b\vec k}^{(L)}=\omega_{\vec k}^{(L)}$, $\tilde\omega_{f\vec k}^{(L)}=\omega_{\vec k}^{(L)}$ in Eq.~(\ref{Eq012b}). We emphasize that the present theory, where $d=2$ explicitly, is not an expansion in the number of dimensions or in large orbital or spin degeneracy but in terms of $n_d$ small. Even the $N\rightarrow\infty$ limit is unessential, used only for the explicit calculation of the coefficients $n_b^{(1)}$, $n_f^{(1)}$, $\tilde\varepsilon_d$, $A_{\vec k,\vec k'}$, $\tilde\omega_{\vec k}^{(L)}$ in Eqs.~(\ref{Eq012a}) and (\ref{Eq012b}).

Apparently, $\GD^{(1)>}$ of Eq.~(\ref{Eq012a}) has some features of the mean-field slave boson theory (MFSBT), namely in Eq.~(\ref{Eq005b}) for $\Gamma_{\vec k}^{(2)}$ it gives the band narrowing and the renormalization of the CT gap, 

\begin{eqnarray}
t_{pd}^2\rightarrow t_{pd}^{(1)2}&=&t_{pd}^2(1+\frac{1}{2}n_b^{(1)})n_f^{(1)}\;,\nonumber\\
\Delta_{pd}\rightarrow\Delta^{(1)}_{pd}&=&\varepsilon_p-\varepsilon_d^{(1)}\nonumber\\&=&\Delta_{pd}-3\beta_\lambda^{(1)}(\omega=\varepsilon_d^{(1)}+\lambda)\;,\label{Eq015zvj}
\end{eqnarray}

\noindent since $\varepsilon_p$, as well as $t_{pp}$, remain unaffected. For $n_d^{(1)}$ small, when the LSR relation $n_b^{(1)}\approx n_d^{(1)}$ holds well according to Eqs.~(\ref{Eq010a}), and $Q^{(1)}=n_f^{(1)}+n_b^{(1)}\approx1$, this reduces to $t_{pd}^2\rightarrow t_{pd}^2(1-n_d^{(1)}/2)$. Such $t_{pd}^2$ renormalization is about half of that predicted\cite{ko1,mr1} by the MFSBT. Concomitantly, and in contrast to the MFSBT, $\Delta_{pd}^{(1)}$ is somewhat increased with respect to $\Delta_{pd}$. It is further interesting to extrapolate Eq.~(\ref{Eq012a}) to larger $\Delta_{pd}$ when $n_d^{(1)}(\Delta_{pd},t_{pd},t_{pp},x)\approx1/2$. $n_d^{(1)}\approx n_b^{(1)}\approx1-n_f^{(1)}$ then gives $t_{pd}^2(1+n_b^{(1)}/2)n_f^{(1)}$ closer to $t_{pd}^2(1-n_d^{MFSB})$ of the MFSBT. The difference here is that MFSBT replaces $n_d^{(1)}$ in Eq.~(\ref{Eq015zvj}) by $n_d^{MFSB}$, which connects it more closely to a self consistent than to the iterative $r =2$ diagrammatic theory. Such differences are small for $n_d^{(1)}$ small and the behavior of Eq.~(\ref{Eq015zvj}) with increasing $n_d^{(1)}$ is satisfactory.

We proceed by taking into account $\GD^{(1)<}$ given by Eq.~(\ref{Eq012b}). In contrast to the coherent band narrowing associated with $\GD^{(1)>}$ of Eq.~(\ref{Eq012a}), the receding $(-i\eta)$ continua of Eq.~(\ref{Eq012b}) describes the dynamic d$^{10}$$\leftrightarrow$d$^9$ CT disorder. Indeed, the convolution (\ref{Eq011}) in frequency means that the transformation of $f^\dagger$-particle (d$^9$ state) in the $b^\dagger$-particle (d$^{10}$ state) is causally correlated in time. In contrast to that, the two $\vec k,\vec k'$ integrations in $\GD^{(1)<}$ are independent, rather than being a convolution in $\vec k,\vec k'=\vec k+\vec q$. The "local" d$^{10}\leftrightarrow d^9$ CT disorder associated with Eq.~(\ref{Eq012b}) is thus the result of a causal temporal and spatial combination of the "local"  slave disorders appearing in "local" $F^{(1)}$ and $B^{(1)}$ which appear for $1+x\neq0$. These correlations are related to the local gauge invariance "to quartic order in $t_{pd}$", which localizes the CT d$^{10}$$\leftrightarrow$d$^9$ event in a single CuO$_2$ cell (it should be however kept in mind that the SFT is not locally gauge invariant unless it is carried to the infinite order in $t_{pd}$). $\GD^{(1)<}$ is thus the first step of the perturbation process which ensures that the $f$-particle is annihilated/created simultaneously with the creation/annihilation of the $b$-particle in the description of the "local" CuO$_2$ CT disorder. In the limit $N\rightarrow\infty$, $\IM\GD^{(1)<}$ is a step-like function, finite in the range $2\mu^{(1)}-\varepsilon_d>\omega>2\omega_M-\varepsilon_d$. This disorder is revealed by the physical single particle propagators $\LD_{\vec k}^{(2)}$ and $t_{pd}^{-2}\Gamma_{\vec k}^{(2)}$ of  Eqs.~(\ref{Eq005a}) and (\ref{Eq005b}), defined by the irreducible self energy $\GD^{(1)}$. Those propagators thus describe in particular the dynamic d$^{10}$$\leftrightarrow$d$^9$ CT disorder of the permanent particles as seen by the intermittent particle.

\begin{figure}[htb]

\begin{center}{\scalebox{0.75}
{\includegraphics{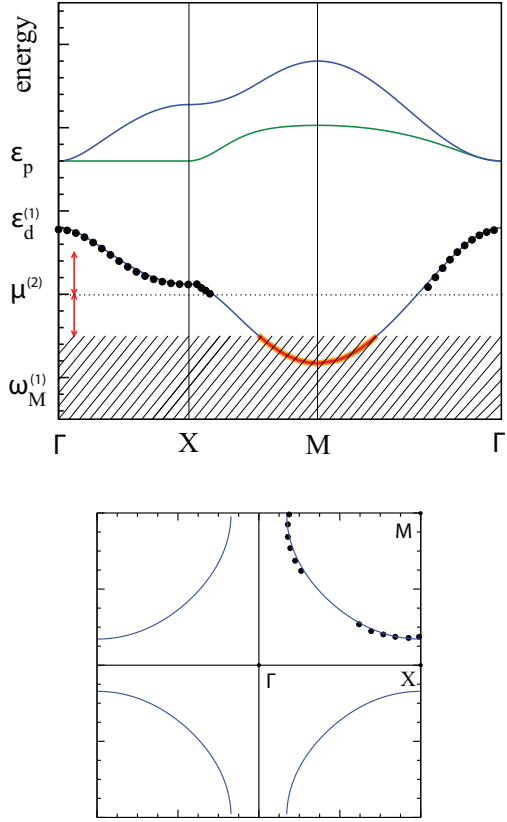}}}
\end{center}

\caption{(Color online) (a) $\IM t_{pd}^{-2}\Gamma_{\vec k}^{(2)}$ spectrum for $\tilde\Delta_{pd}/|t_{pp}|=8/3$, $\tilde t_{pd}/|t_{pp}|=5/3$, $t_{pp}= -0.3$ eV.  Experimental\cite{yo1} points correspond to $x=0.07$ LSCO; (b) $\mu^{(2)}$  is taken in Sec.~\ref{Sec11.1} to satisfy the Luttinger sum rule. The leading d$^{10}\leftrightarrow$d$^9$ disorder effects are included schematically by gray and red shadings; (b) Characteristic $\omega=\mu^{(2)}$ Fermi arcs obtained for $0<x<x_{vH}$  correspond to the $\pi/4$ rotation of the Fermi surface induced by  $t_{pp}$.\label{fig004}}
\end{figure}

Let us thus consider $\Gamma_{\vec k}^{(2)}$ of Eq.~(\ref{Eq005b}) in some more detail, keeping in mind that a parallel discussion exists for $D_{\vec k}^{(2)}$. The result is particularly transparent in the approximation which takes into account that for $n_d^{(1)}$ small $\IM\GD^{(1)<}$ is small with respect to the energy distance between the affected pole at $\omega_{\vec k}^{(L)}$ and the Fermi level. In this, "quasiparticle" limit $\RE\GD^{(1)<}$ is unimportant. The poles and the associated residue are then given by the coherent band narrowing $\omega_{j,\vec k}^{(j)}$ of the three bands $\omega_{\vec k}^{(j)}$, shown in Fig.~\ref{fig004}, which is given essentially by replacing $\GD^{(0)}$ by $\GD^{(1)>}$ in $\Gamma_{\vec k}^{(1)}$. $z_{L,\vec k}^{(1)}$ are the corresponding residues, such as those of Eq.~(\ref{Eq008}). $\IM\GD^{(1)<}$ is then the only contribution to $\GD^{(1)}$ beyond the coherently renormalized hybridization. In the relevant frequency range this gives
 
\begin{equation}
\IM t_{pd}^{-2}\Gamma_{\vec k}^{(2)}(\omega)\approx\sum_j\frac{z^{(1)}_{j,\vec k} (\eta_S+t_{pd}^2\IM\GD^{(1)<})}{(\omega-\omega_{j,\vec k}^{(1)})^2+(\eta_S+t_{pd}^2\IM\GD^{(1)<})^2}\label{Eq012c}\;.
\end{equation}

\noindent where $\eta_S=\eta\sign(\mu^{(2)}-\omega_{j,\vec k}^{(1)})$ and $\IM\GD^{(1)<}$ is taken at $\omega=\omega_{L,\vec k}^{(1)}$.

The step-like continuum in $\IM\GD^{(1)<}$ of Eqs.~(\ref{Eq012b}) and (\ref{Eq012c}) falls in the range $2\omega_M-\varepsilon_d<\omega<2\mu^{(1)}-\varepsilon_d$, where $\omega_M$ denotes the bottom of the $L$-band,\cite{mr1} i.e. the disorder is far from the Fermi level. $\GD^{(1)<}$ is a receding term and, as a consequence, $\IM\GD^{(1)<}$ carries the same sign as $\eta_S$. This contrasts with the $F_\lambda$-spinless fermion where the appearance of the disorder is related to the double change in sign of the imaginary part of the self energy (see Appendix) and so to the annihilation of $f$-fermions. Note that the sign of $\IM\GD^{(1)}$ could not be determined from the fact that the step-like continuum falls below the Fermi level but rather, its receding $(-i\eta)$ nature had to be demonstrated directly in Eq.~(\ref{A10}) from its position in the upper half-plane. This illustrates the importance of the causal time-dependent analysis, carried out here. According to this analysis the continuum in $\IM\GD^{(1)<}$ is to be interpreted as the lifetime effect of the coherently hybridized occupied states, 
      
\begin{equation}
\frac{2\pi}{\tau_{pd}(\omega^{(1)}_{L,\vec k})}=t^2_{pd}\IM\Sigma^{(1)<}(\omega_{L,\vec k}^{(1)})\label{Eq019zvj}
\end{equation}

\noindent rather than as related to the average $p-d$ CT. In the quasiparticle approximation the effect of $\IM \GD^{(1)<}$ is thus a broadening of the affected Dirac functions into the Lorentzians normalized to unity, shown by the broad line in Fig.~\ref{fig004}.

One should however take into account that, rather than zero, $\IM\GD^{(1)<}$ of Eq.~(\ref{A10}) is finite {\it around} the $L^{(1)}$-band, all over the Brillouin zone in the frequency range $2\omega_M-\varepsilon_d<\omega<2\mu^{(1)}-\varepsilon_d$, where the coherent hybridization is absent. The coherent and incoherent frequency ranges contribute essentially additively to average occupations of the O (and Cu) sites. The conservation of the quasi particle spectral density in Eq.~(\ref{Eq019zvj}) means in particular that $\IM \GD^{(1)<}$ does not affect the contribution $n_p^{(2HF)}$ of the extended states to $n_p^{(2)}=n_p^{(2HF)}+n_p^{(inc)}$ for a given chemical potential $\mu^{(2)}$. Furthermore, by using Eq.~(\ref{Eq005a}) for $\LD_{\vec k}^{(2)}$ either in the Bethe-Salpeter or in the Dyson form we can easily convince ourselves from Eqs.~(\ref{Eq005a}) and (\ref{Eq012a}) that, in addition to 
      
\begin{equation}
n_p^{(2)}\approx n_p^{(2HF)}+ n_p^{(2inc)}\;,
\end{equation}
          
\noindent we have                    
                      
\begin{equation}
n_d^{(2)}\approx(1+\frac{1}{2}n_b^{(1)})n_f^{(1)}n_d^{(2HF)}+ n_d^{(2inc)}\;, \label{Eq013a}
\end{equation}            		              

\noindent where $n_p^{(2inc)}$ and $n_d^{(2inc)}$ describe the incoherent contributions of $\IM \GD^{(1)<}$. In the simplest approximation $n_d^{(2inc)}$ is obtained by integrating $\IM \GD^{(1)<}$ of Eqs.~(\ref{Eq012b}) and (\ref{A10}) over $\omega$, while $n_p^{(2inc)}$ is neglected. The first term in $n_d^{ (2)}$ is somewhat smaller than $n_d^{(2HF)}$ calculated from the HF bands defined by the self energy $\IM \GD^{(1)>}$ divided by $(1+n_b^{(1)}/2)n_f^{(1)}$, while the second term increases $n_d^{(2)}$. The chemical potential $\mu^{(2)}$ is finally determined from the sum rule $n_d^{(2)}+2n_p^{(2)}=1+x$. Equation (\ref{Eq013a}) combined with Eq.~(\ref{Eq015zvj}) establishes so the connection between the $r =2$ expansion and the familiar HF theory.
  
For  $n_d^{(1)}\approx n_b^{(1)}\approx1-n_f^{(1)}$ small Eq.~(\ref{Eq013a}) can be used, on linearizing in terms of  $n_d^{(2H)}-n_d^{(1)}$, to write 

\begin{eqnarray}                              
&&n_d^{(2H)}+2n_p^{(2H)}\approx1+x_{eff}\;,\nonumber\\
&&x_{eff}=x+\frac{1}{2}n_d^{(1)2}-(n_d^{(2inc)}+2n_p^{(2inc)})\label{Eq013b}\;.
\end{eqnarray}
                                                                                            
\noindent $\mu^{(2)}$ corresponds thus to the HF chemical potential for the effective doping $x_{eff}$ put into the $L^{(1)}$-band renormalized according to Eq. ~(\ref{Eq015zvj}). $n_d^{(2H)}$ can be then found straightforwardly, using the conventional HF procedure described briefly in Sec. \ref{Sec04}. Notably, Eq.~(\ref{Eq013b}) explicitly breaks the standard Luttinger sum rule (to be distinguished from the LSR $n_d=n_b$). Apparently, the shift $\mu^{(2)}-\mu^{(1)}$ which corresponds to the shift $n_d^{(2H)}-n_d^{(1)}$ is small for $n_d^{(1)}$ small. Indeed, the large shift of the chemical potential $\mu^{(1)}-\mu^{(0)}$ due to the anticrossing of the d-state and two p-bands, obtained under assumption that $n_d^{(1)}-n_d^{(0)}$ ($n_d^{(0)}=0$) is small, was achieved in the first $r=1$ step of the present calculation. 
       
The above discussion shows in particular that the regime of $n_d^{(1)}$ small corresponds to a positive $\partial n_d^{(2)}/\partial x\approx\partial n_d^{(2H)}/\partial x$  at $x$ small. In other words, the crossover for x small is absent at $r=2$ in the parametric regime $\Delta_{pd}^{(1)2}\approx2t_{pd}^{(1)2}\approx\Delta_{pd}^{(1)}|t_{pp}|\geq4t_{pp}^2$. The condition $\partial n_d^{(2H)}/\partial x\approx0$ with $n_d^{(2H)}$ of Eq.~(\ref{Eq013b}) bears some resemblance with the corresponding MFSB calculation. As already mentioned, $\partial n_d^{(MFSB)}/\partial x=0$ at fixed $\Delta_{pd}/t_{pp}$, $t_{pd}/t_{pp}$ was shown\cite{ku1} to occur at small $_{cs}^{MFSB}$ for appreciable $n_d^{(MFSB)}$ (e.g.\cite{ku1} for $\Delta_{pd}\approx4t_{pd}\gg|t_{pp}|$ when $n_d^{(1)}\approx1/2$ and $n_d^{(MFSB)}\approx3/4$). This indicates that $\partial n_d^{(r)}/\partial x\approx0$  might also be achieved at the $r =2$ level. However, expectedly, $\mu^{(2)}-\mu^{(1)}$ becomes appreciable for $n_d^{(2)}$ close to $3/4$. Such large variations of $\mu^{(r)}$ with $r$ signify that the absolute convergence of the present theory becomes poor close to the crossover. The precise numerical analysis of the condition $\partial n_d^{(2)}/\partial x\approx0$ would thus be unreliable and one can use instead $n_d^{(1)}\leq1/2$, $n_d^{(2)}\leq2/3$ as a reasonable estimate of $x_{cs}^{(2)}$ from the $x\geq x_{cs}^{(2)}$ side.

It should be finally pointed out that $\LD_{\vec k}^{(2)}$ and $\Gamma_{\vec k}^{(2)}$ given by Eqs.~(\ref{Eq012a},\ref{Eq012b},\ref{Eq012c}) are not sensitive to the omission of the Cu-O (anti)commutation rules. $\GD^{(1)}$ of Eq.~(\ref{Eq011}) is determined by $B_\lambda^{(1)}$ and $F_\lambda^{(1)}$ which in turn are determined by $\beta_\lambda^{(1)}$ and its $f\leftrightarrow b$ symmetric counterpart $\phi_\lambda^{(1)}$. The latter are determined by $\Gamma_{\vec k}^{(1)}$, which is insensitive to the Cu-O commutation and, therefore, so are $\LD_{\vec k}^{(2)}$ and $\Gamma_{\vec k}^{(2)}$. This line of reasoning can be extended even further by assuming that $\GD^{(r)}\sim B_\lambda^{(r)}\star F_{\lambda}^{(r)}$, which is named\cite{ni1} the $U_d=\infty$ non crossing approximation (NCA). Since the NCA is $f\leftrightarrow b$ symmetric it is $\lambda$-independent and satisfies $Q_{\vec R}\approx1$ "on average" for $n_d$ small. It can therefore be called a "conserving" approximation.\cite{kr1} According to previous numerical slave boson NCA\cite{ni1} and finite $U_d$ DMFT calculations\cite{zl1} such corrections are expected to spread the d$^{10}$$\leftrightarrow$d$^9$ disorder effects all over the coherent spectrum. However, these calculations\cite{ni1} start from the $n_d^{(0)}=1$ HF ground state of $b$-fermions which is not locally gauge invariant, in contrast to the $n_d^{(0)}=0$ ground state of Eq.~(\ref{Eq001}) and those results should be therefore taken with some caution. Although Eqs.~(\ref{Eq012b},\ref{Eq012c}) show that the dynamical d$^{10}$$\leftrightarrow$d$^9$ CT disorder induced by $t_{pd}$ is an inherent property of the large $U_d$ Emery model in the regime $\Delta_{pd}\geq4|t_{pp}|$ under consideration, further analysis is required to determine the full temporal $(\omega\pm i\eta)$ structure of the higher order CT terms, before comparison with experiments. Most importantly, the present analysis indicates that the intrinsic dynamic d$^{10}$$\leftrightarrow$d$^9$ charge disorder effect is "expandable" in terms of $n_d$ small, i.e. that it does not affect the Emery particle propagation strongly in the vicinity of the Fermi level, provided that $n_d$ is small. It should be kept in mind however that a strong electron-phonon coupling, omitted here, can slow down\cite{go1} the d$^{10}$$\leftrightarrow$d$^9$ CT disorder and bring its effects to the Fermi surface.

\section{Vertex corrections to single particle \texorpdfstring{$p$}{p}- and \texorpdfstring{$d$}{d}-propagators}

\begin{figure}[htb]

\begin{center}{\scalebox{0.5}
{\includegraphics{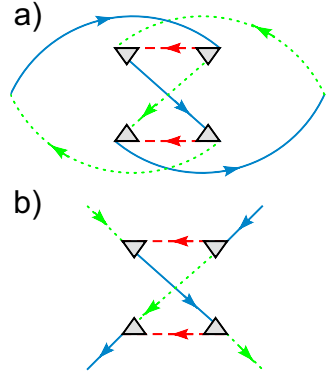}}}
\end{center}

\caption{(Color online) Skeleton contribution (a) to $\GD^{(2)}$ generated by effective two-particle $b$-$f$ interaction (b).\label{fig005}}

\end{figure}

The lowest order diagram for $\GD$ that is not taken into account by the NCA is shown in Fig.~\ref{fig005}. As in previous low order diagrams, the arrows of time are carefully included to show the flows of energy throughout the diagram. Fig.~\ref{fig005} gives a local vertex correction to $\GD^{(2)}$ related to the local effective $b-f$ interaction. It is clear that when the anticrossing of p-bands and the d-state is important the $p$-propagators in Fig.~\ref{fig005} have to be interpreted as the $pdp$ hybridized propagators $t_{pd}^{-2}\Gamma_{\vec k}^{(1)}$ in the strictly iterative scheme. Such diagrams take into account that two permanent $p$-holes cannot hop {\it simultaneously} to a given Cu-site, one by the creation of the $b$-boson and the other by the annihilation of the spinless fermion because the Cu-site carries either boson or a spinless fermion at any one time. Those contributions are allowed in the NCA, and thus have to be subtracted from Eq.~(\ref{Eq012b}). However, when $n_d$ is small the number of unwanted coincidences is expectedly much smaller than the total number of incoherent processes and this correction will therefore be ignored here. 

\begin{figure}[tb]

\begin{center}{\scalebox{0.5}
{\includegraphics{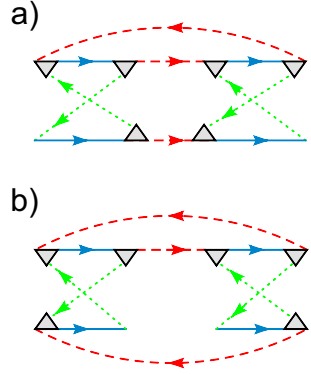}}}
\end{center}

\caption{(Color online) Skeleton diagrams which contribute to irreducible $\GD_{\vec k}^{(3)}$: (a) through the particle-particle channel; (b) through the particle-hole channel.\label{fig006}}

\end{figure}

Further on, the two "squares" $\Lambda_0^b$ in Fig.~\ref{fig006} which involve $b$ and $f$ propagators cannot be reduced to the self energy renormalizations of one of the propagators $B^{(0)}$ or $F^{(0)}$, which appear in the diagrams of Fig.~\ref{fig006}. The latter thus also extends beyond the NCA. $\GD^{(3)}$ depends on $\vec k$ and so it contains a nonlocal component. In this respect, we emphasize the appearance in Fig.~\ref{fig006} of two particle-particle or particle-hole $pdp$-lines connecting two $t_{pd}^4\Lambda_0^b$'s. In the SFT, the single particle lines can carry arbitrary spins.  The corresponding elementary bubbles open singular particle-hole and the particle-particle correlation channels. This phenomenon is the seed of the pseudogap effects of the magnetic (SDW), charge/bond (CDW/BOW), and Cooper pairing correlations within $\GD_{\vec k}^{(3)}$. Those terms and their extension to higher orders are therefore of utmost physical importance, in addition to the nonmagnetic $r=1,2,3$ terms describing in particular the local, dynamic d$^{10}$$\leftrightarrow$d$^9$ CT disorder. Apparently, the explicit nonlocality of $\GD_{\vec k}^{(3)}$ is useful for comparison with the NCA and the DMFT results.\cite{zl1,mc1,rb1}

\begin{figure}[htb]

\begin{center}{\scalebox{0.5}
{\includegraphics{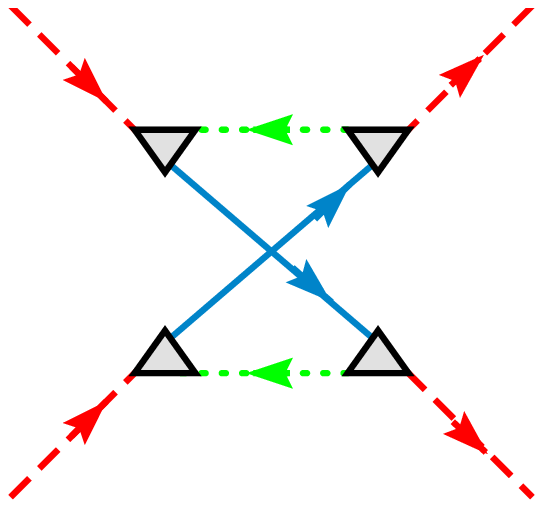}}}
\end{center}

\begin{center}{\scalebox{0.5}
{\includegraphics{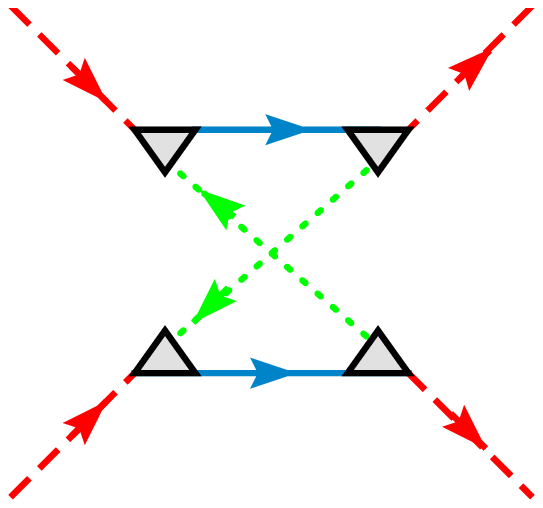}}}
\end{center}

\caption{(Color online) Skeleton diagrams for two $b\leftrightarrow f$ symmetric  effective $p$-particle interactions.\label{fig007}}

\end{figure}

The terms for $\GD_{\vec k}^{(3)}$  additional to those shown in Fig.~\ref{fig006} are easily found on noting that the $F_\lambda^{(0)}$ and $B_\lambda^{(0)}$  propagators must occur $f\leftrightarrow b$ symmetrically in $\GD_{\vec k}^{(3)}$, in order to preserve the local gauge invariance. This corresponds to the replacement of $\Lambda_0^b$ in Fig.~\ref{fig006} by $\Lambda_0^a +\Lambda_0^b$ (independent of $\lambda$) for each spin, shown in Fig.~\ref{fig007}. The four-leg vertices shown in Fig.~\ref{fig007} describe the effective interactions between the $pdp$ hybridized particles. When the $\vec k$-dependences of the triangular vertices are associated with the $pdp$ propagating lines, e.g. with $t_{pd}^{-2}\Gamma_{\vec k}^{(1)}$, the effective interaction between the $pdp$ propagators is local ($t_{pd}^4\Lambda_0$), where $\Lambda_0^{a,b}$  are the convolutions of two $B_\lambda^{(0)}$ and two $F_\lambda^{(0)}$ propagators, e.g.

\begin{equation}
t_{pd}^4\Lambda^b_0=t_{pd}^4\frac{2\varepsilon_d-\omega_1-\omega_2}
{\prod_s(\varepsilon_d-\omega_s-2i\eta)}\label{Eq014}\;.
\end{equation}

\noindent Here, the $\omega_s$ denote the external frequencies which run counterclockwise around $\Lambda_0^b$ in Fig.~\ref{fig005}, with $\omega_1$ in the upper left corner, so that $\omega_1+\omega_2=\omega_3+\omega_4$. Remarkably, $t_{pd}^4\Lambda_0^b$, independent of $\lambda$, is the product of four poles. Fig.~\ref{fig007} show that $t_{pd}^4\Lambda_0^{a,b}$ are kinematic interactions, in the sense that one particle has to wait for the other to leave a given Cu site before it hops to this site. The causal nature of this result is associated with the $-2i\eta$ position of the poles in Eq.~(\ref{Eq014}). 

Analogous kinematic terms were first invoked\cite{ka1} in the $T=0$ (single band) theory of transition metals with sizeable $U_d$ and recently extended\cite{fr3} to high T$_c$ cuprates in the single band limit. Eq.~(\ref{Eq014}) thus represents the generalization\cite{bb1} of this concept to the multiband Emery model in the small $\Delta_{pd}$, $n_d^{(0)}=0$, $U_d=\infty$ limit. It should be noted in this respect that, in the regime $n_d^{(1)}<1/2$, the effective interaction $t_{pd}^4\Lambda_0$  is reasonably small itself. Indeed, upon assuming it to be small, one affects only the hybridized states on the Fermi level. Consistently, the important values of $t_{pd}^4\Lambda_0$, small to moderate, are of the order of

\begin{equation}
t_{pd}^4\Lambda_0\sim\frac{t_{pd}^4}{\Delta^3_{d\mu}}<t_{pd}\label{Eq015}\;.
\end{equation}

\noindent The reason is that, in the $\Delta_{pd}^2\gtrsim2t_{pd}^2\gtrsim\Delta_{pd}|t_{pp}|\gtrsim4t^2_{pp}$ regime $\Delta_{d\mu}=\varepsilon_d-\mu^{(1)}>t_{pd}$ for $\mu^{(1)}<\omega_{vH}\; (\varepsilon_d-\omega_{vH}$ depends only on $t_{pd}$ and $\Delta_{pd}$ and, e.g., is linear in $t_{pd}$ for small $\Delta_{pd}$). The SFT theory therefore converges quickly towards approximate local gauge invariance even when the processes of Fig.~\ref{fig006} are taken into account. 

\begin{figure}[htb]

\begin{center}{\scalebox{0.5}
{\includegraphics{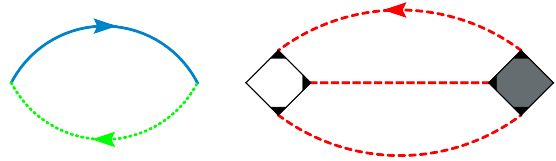}}}
\end{center}

\caption{(Color online) An infinite order expression for $\GD_{\vec k}(\omega)$; the first local term corresponds to the NCA.\label{fig008}}

\end{figure}

Actually, the diagrams shown in Fig.~\ref{fig006}, combined with Fig.~\ref{fig007}, are the only skeleton diagrams for $\GD_{\vec k}$ based only on the vertex $t_{pd}^4\Lambda_0$. Assuming that the effect of $t_{pd}^4\Lambda_0$ on $\GD_{\vec k}$ is dominant (e.g. enhanced by nesting\cite{dz2,bb1,sk3}) it is of some interest to construct the corresponding $r\rightarrow\infty$ $\lambda$-independent expression for $\GD_{\vec k}^{(\infty)}$ on neglecting the off-diagonal $b-f$ interactions of Fig.~\ref{fig005}. The full infinite order partial sum, independent of $\lambda$, is formulated in terms of the renormalized value of $\Lambda_0$, $\Lambda$ and $\Gamma_{\vec k}$ as follows. First, one introduces the infinite order slave-particle Dyson propagators in $\Lambda_0$ instead of the bare ones, denoting the result by $\Lambda_0$. In Fig.~\ref{fig008} this is depicted by the open square. Second, as suggested by Fig.~\ref{fig006} and analogously to the usual single band perturbation theory, one joins $\Lambda_0$ with the fully renormalized four-leg vertex $\Lambda$, shaded square, using the $pdp$-bubble formed from two $t_{pd}^{-2}\Gamma_{\vec k}$'s, denoted as arrowless in the figure. Convoluting the result with $t_{pd}^{-2}\Gamma_{\vec k}$ one finally obtains $\GD_{\vec k}(\omega)$, which is depicted symbolically in Fig.~\ref{fig008}. It is noteworthy that the nonlocal nature of $\GD_{\vec k}$ then corresponds to the nonlocal nature of $t_{pd}^{-2}\Gamma_{\vec k}$. The backward arrow in the second term of Fig.~\ref{fig008} is shown in order to stress that, in generalizing the skeleton diagram of Fig.~\ref{fig006}, there must be a $pdp$-line which starts and finishes in Fig.~\ref{fig008} with receding $\Gamma_{\vec k}^{(0)<}$ of Eq.~(\ref{Eq006}), so involving the occupied $\vec k$'s. The second term in Fig.~\ref{fig008} thus vanishes for $1+x=0$ while the first reduces to $\GD^{(0)}=\LD^{(0)}$, as the exact $\GD_{\vec k}$ should. On the other hand, for $1+x=0$ the vertex $t_{pd}^4\Lambda_0$ of Fig.~\ref{fig007} remains finite, in contrast to that of Fig.~\ref{fig005}b. The approximation of Fig.~\ref{fig008} is thus nearly exact in the low density (electron doping) limit. This will be taken up again in Sec.~\ref{Sec08} while, in the following, we continue to discuss the approximation of Fig.~\ref{fig008} with the $n_d$ small in mind.

\section{Spin disorder and RVB fluctuations}

\begin{figure}[htb]

\begin{center}{\scalebox{0.75}
{\includegraphics{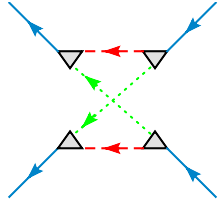}}}
\end{center}

\caption{(Color online) Skeleton diagram for $b-b$ interaction, which generates the superexchange.\label{fig009}}

\end{figure}

Figure~\ref{fig008} for $\GD_{\vec k}(\omega)$ omits the $b-f$ interaction of Fig.~\ref{fig005} but not necessarily the "diagonal" $b-b$ superexchange interactions stemming from Fig.~\ref{fig009} which can be partially included in Fig.~\ref{fig008} via the renormalization of single $b$-particle propagators. Those effects, as well as the effects beyond Fig.~\ref{fig008}, are conveniently discussed using the spin-flip correlation function $\chi_{bb}^{\uparrow\downarrow}(\vec q,t)=-(i/N)\langle\hat T\sum_{\vec k}b_{\vec k,\uparrow}^\dagger b_{\vec k+\vec q,\downarrow}b_{\vec k,\uparrow}(t)b_{\vec k+\vec q,\downarrow}^\dagger(t)\rangle$ also related to the spin flip $dpd-dpd$ correlation function $\chi_{dd}^{\uparrow\downarrow}(\vec q,t)$. Indeed, it is long known\cite{tu1,fu1} from the slave boson theories that the super exchange interactions appear as interactions between two auxiliary spin carriers $b_\sigma$. The reason is that the spin-flip operator on the Cu-site $s^\dagger= c_\uparrow^\dagger c_\downarrow$ maps on $b_\uparrow^\dagger b_\downarrow$ in the slave particle theories. It means that two holes of opposite spins on an intermediate O-site can hop simultaneously, each to one of two empty Cu neighbors. Notably, the effective interaction of Fig.~\ref{fig009} is purely antiferromagnetic if  $t_{pd}^{-2}\Gamma_{\vec k}^{(0)<}$ with $t_{pp}=0$ is used, otherwise the effective interactions between next-to-next-Cu neighbors are generated. The internal square in Fig.~\ref{fig009} is thus, quite schematically, of the order of the squared number of occupied $pdp$ states, $(1+x-n_d)^2$. Although such an interaction is appreciable for $|x|$ small, the zeroth-order particle-hole correlation $\chi_{bb}^{(0)}$ among the auxiliary spin carriers vanishes i.e. $\chi_{bb}^{(0)}$ with $B^{(0)}=B^{(0)>}$ of Eq.~(\ref{Eq004}). The external lines in Fig.~\ref{fig009} should be therefore associated with $B^{(1)}=B^{(1)>}+B^{(1)<}$ of Eq.~(\ref{A3}). The corresponding elementary correlation function, proportional to $B^{(1)}\ast B^{(1)}$,

\begin{equation}
\chi_{bb}^{(1)}(\omega)=\frac{1}{N}\sum_{\vec k}\frac{(1+\frac{1}{2}n_b^{(1)})\gamma_{\vec k}^{(b)}f_{\vec k}^{(L)}}{\omega-\omega^{(L)}(\vec k)+\varepsilon_{db}^{(1)}+3i\eta}\;,\label{Eq022zvj}
\end{equation} 
                                                                                              
\noindent is therefore linear in $n_b^{(1)}\approx n_d^{(1)}$ to the leading order,  i.e. negligible for $n_d^{(1)}$ small, compared to the corresponding correlations $\chi_{pp}^{(1)}$ between the $pdp$-propagators, discussed in Sec.~\ref{Sec10}. Moreover, in sharp contrast with $\chi_{pp}^{(1)}$, $\chi_{bb}^{(1)}$ is dispersionless and incoherent in the range $\mu^{(1)}-\varepsilon_{db}^{(1)}\geq\omega\geq\omega_M-\varepsilon_{db}^{(1)}$, i.e., $\chi_{bb}^{(1)}(\omega)$ is real around $\omega=0$. Notably, the spin-flip $\chi_{bb}^{(1)}$ describes the local dynamic spin disorder on the Cu-sites and is the only one renormalized further by effective interactions. 

Indeed, the analog of the genuine superexchange interaction $J_{pd}(\vec q,\omega)$ (i.e. RKKY $J$, $J'$, $J''$) of the permanent spins\cite{ge1,za1,tu1,fu1,al1} on the Cu-sites appears (together with mixed terms) when the internal spinless fermion lines in Fig.~\ref{fig009} are taken as  $F^{(1)}=F^{(1)<}+F^{(1)>}$ ($f\leftrightarrow b$ symmetrically with Eq.~(\ref{A3})) rather than as $F^{(0)}=F^{(0)<}$, the empty $pdp$ states are then involved too. This interaction itself is thus schematically given in powers of $n_d^{(1)}$, i.e for $n_d^{(1)}$ small it is small with respect to the effective interaction $t_{pd}^4\Lambda_0$ between the $pdp$-propagators (not surprisingly, just the contrary was suggested\cite{fu1} upon coming from the $x<x_{cs}$, $n_d^{(0)}=1$, Mott side). However again, in sharp contrast to $t_{pd}^4\Lambda_0$, the superexchange interaction involves the $F^{(1)>}$ disorder, $f\leftrightarrow b$ symetrically with $\chi_{bb}^{(1)}$. In the small $n_d$ limit the overall effects of the magnetic {\it interactions} between the $b$-particle-hole propagators come thus out as small and incoherent with respect to those of the interaction $t_{pd}^4\Lambda_0$ between the $pdp$ particle-hole propagators. Notably, $\chi_{bb}^{\uparrow\downarrow}(\vec q,\omega)$ resulting from the convolutions $\chi_{bb}^{(1)}(\omega)\ast J_{pd}(\vec q,\omega)\ast\chi_{bb}^{(1)}(\omega)$  can then be interpreted as describing a dilute gas of RVB singlets\cite{an1,an2} within the uncorrelated {\it spin}  disorder of Eq.~(\ref{Eq022zvj}).
      
It is appropriate to add here some qualitative considerations about what can be expected when $n_d^{(2)}$ approaches $1/2$ for $x\geq x_{cs}$ which makes $J_{pd}$ and $t_{pd}^4\Lambda_0$ comparable. This corresponds first to the evolution from the gas to the liquid of RVB singlets. Next we note that incoherencies of $\chi_{bb}^{(1)}(\omega)$ and $J_{pd}(\vec q,\omega)$ are related to the dispersion of the $L$-band. In higher order calculations required for $x\approx x_{cs}$ the local AF RVB are likely to progressively localize the $pdp$-states within the CuO$_2$ unit cell, i.e. it becomes better to take empty $pdp$ states as nearly dispersionless. In particular, intercell $t_{pp}$ should then be practically renormalized out,\cite{va2} which increases $n_d$ with respect to $n_d^{(2)}$. The resulting Mott-AF coherence of the RVB singlets amounts to their short range AF ordering. This in turn plays in favor of increasing the number of empty $pdp$ states i.e. tends to increase the spin polarized $n_d$ further. On balancing such Mott-AF effects against the steady decrease of $n_d^{(2)}$ with decreasing $x$ towards $x_{cs}$ from above one thus expects so to reach the saturation $\partial n_d/\partial x=0$ at small $x\approx x_{cs}$. Such an expectation is corroborated by the $x<x_{cs}$ slave boson calculation\cite{bbk} which starts perturbatively from the $n_d^{(0)}=1$ Ne\' el state and shows that $n_d$ in the $x=0$ Mott-AF phase is in low orders larger than $n_d$ in the small $x>x_{cs}$ metallic phase, all parameters of the Emery model except small $x$ being kept equal.

Although this kind of analysis may be extended in principle to evaluate the magnetic correction to $x_{cs}^{(2)}$ we limit ourselves here to the observation that the low order SFT is consistent for $x>x_{cs}$ even when $n_d$ is somewhat larger than $n_d^{(2)}$.

\section{Modified slave fermion theory (MSFT)\label{Sec08}}

The present approach derives the effective $pdp$-$pdp$ interaction $t_{pd}^4\Lambda_0$ in the small to moderate $\Delta_{pd}$ regime, where it represents the dominant effective interaction. Its $T=0$ causal properties are explicitly related to local gauge invariance. Its shortcoming is, however, that it does not take into account the Cu-O anticommutations. This is reflected in the spin structure of the effective interaction $t_{pd}^4\Lambda_0$ of Figs.~\ref{fig007}a,b. Due to spin conservation in the triangular vertices of Fig.~\ref{fig007}, the effective interaction $t_{pd}^4\Lambda_0$ of the hybridized $t_{pd}^{-2}\Gamma_{\vec k}^{(1)}$ particles is the same in the singlet as in the triplet channel. Thus the assistance of slave fermion particles is required to forbid the double occupation of the Cu-site in both spin channels. However the triplet scattering in the original Hamiltonian is expected to vanish identically, irrespective of the value of $U_d$ ($U_d=\infty$ included) due to Cu-O anticommutations. It is in this aspect that the theory which accounts for the Cu-O anticommutations {\it a posteriori}, requires a modification of the SFT.

We start by emphasizing that adding intermittently the first $p$- or $d$-particle ($1+x=0$) has to lead to the coherent single-particle hybridization irrespective of the value of $U_d$ and the anticommutation rules, as was discussed in connection with Fig.~\ref{fig008}. The SFT reproduces this result by reducing $\GD_{\vec k}$ to $\GD_0$, i.e. $\Gamma_{\vec k}=\Gamma_{\vec k}^{(1)}$ and $\LD_{\vec k}=\LD_{\vec k}^{(1)}$, i.e. only the coherent $p-d$ hybridization is retained (irrespective of whether the anticrossing is important or not). Likewise, one finds $B=B^{(0)}$ and $F=F^{(0)}$, in a manner analogous  to the phonon propagator in the standard polaron problems.\cite{ob1}
      
Next we turn to the $N(1+x)=1$ "bipolaron" case. This is the fundamental problem in which the two requirements of no double occupancy and the $d-p$ anticommutation rules come into play simultaneously. There are two ways to approach the problem of two particles in the system. In one approach, a single $N(1+x)=1$ fermion, say with spin $\uparrow$, is placed in the $\vec k=[\pi,\pi]/a$  state of the $L$-band and the SFT is applied to the propagation of the additional $p$ (or $d$) particle with spin either $\uparrow$ or $\downarrow$. This is covered in the SFT by the slight adaptation to the fact that the spin in the ground state is now unpaired (time reversal symmetry is broken). The alternative approach involves the simultaneous creation, on the top of the empty p-band, of two particles in the singlet or in the triplet state. Obviously both approaches must coincide in giving the essence of the "bipolaron" physics.

Let us thus start by considering the cohabitation of two $p$-particles in the SFT as a scattering problem. The fact that singlet and triplet scatterings of Fig.~\ref{fig007} are equal in the SFT bears some resemblance to the text-book scattering\cite{msh} of two distinguishable particles of spin $1/2$ by a local interaction. We see, therefore, that by treating the $f$-, $b$- and $p$-particles as distinguishable, ignoring consequently the $d$-$p$ anticommutation rules, the SFT forbids double occupancy for both triplet and singlet scattering equally and dependently on band parameters $\varepsilon_d$ and $t_{pd}$.

On the other hand, with the original Emery $U_d$ Hamiltonian it is the Pauli principle which forbids two particles in the triplet configuration to occur simultaneously on the Cu-site. The triplet scattering by the local interaction and its effects therefore vanish identically. By contrast, the scattering of two particles with opposite spins by $U_d$ is finite and in the $U_d=\infty$ limit has to correspond to the effect of forbidding the simultaneous double occupancy of the Cu site. In the conventional strong $U_d$ coupling multiband perturbation theory only this effect can give rise to the effective singlet repulsion of two $p$-fermions independent of $U_d$ but depending on the band parameters, $t_{pd}$ in particular. 

Returning to the slave-particle formulations this means that, in the MSFT, we have to remove by hand all diagrams in which the two internal $b$-lines in vertices of Fig.~\ref{fig007} carry the same spin projection and keep only those with two opposite projections. $t_{pd}^4\Lambda_0$ of the diagrams in Fig.~\ref{fig007} then represents the $n_d^{(0)}=0$ counterpart of the $n_d^{(0)}=1$ result\cite{ge1,za1,tu1,fu1} for the superexchange $J_{pd}=4t_{pd}^4/\Delta_{pd}^3$ in the $U_d=\infty$ limit for $\Delta_{pd}$ large. In the standard scattering problem\cite{msh} with a local interaction, the {\it a posteriori} antisymmetrization of the wave function of two spin $1/2$ particles makes the triplet scattering amplitude vanish, simultaneously doubling the unsymmetrized singlet scattering amplitude. In the present case, Fig.~\ref{fig007} describes, in second quantization, the singlet scattering of two $p$-particles, which are not completely distinguishable because anticommuting between O-sites is taken into account. The problem therefore requires additional care.

\begin{figure}[htb]

\begin{center}{\scalebox{0.5}
{\includegraphics{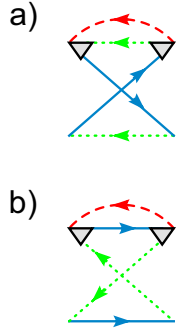}}}
\end{center}

\caption{(Color online) Irreducible $b\leftrightarrow f$ symmetric renormalizations for $\GD^{(1)}$ quadratic in $t_{pd}$ involving: (a) the local renormalization of the $b$-propagator and (b) the local renormalization of the $f$-propagator.\label{fig010}}

\end{figure}

In determining the normalization factors of the scattering terms in the MSFT it is important to note that the SFT/MSFT self energy corrections, which amount to treating one particle in the "bipolaron" problem as permanent, are obtained by closing one of the two $p$-lines of Figs.~\ref{fig007}a,b. Actually, rather than dwelling upon the "bipolaron" case, where the MSFT is to be exact, we can immediately address the limit of $n_d$ small. This reflects the fact that the necessary and sufficient condition for the fast convergence of the present theory is that $n_d$ is small, which is of course fulfilled, especially for $1+x$ small (in this latter case for arbitrary band parameters). The resulting $\Delta\GD^{(1)}$ is shown in Fig.~\ref{fig010}. Clearly it corresponds to the leading contributions to $\GD^{(1)}$ of Eq.~(\ref{Eq011}) given by   

\begin{equation}
\Delta\GD^{(1a)>}+\Delta\GD^{(1b)<}=\frac{-i}{2\pi}(\Delta B_\lambda^{(1)}\ast F_\lambda^<+B_\lambda^>\ast\Delta F_\lambda^{(1)})\label{Eq016}\;,
\end{equation}

\noindent where $\Delta B_\lambda^{(1)}=(B_\lambda^{(0)})^2\beta_\lambda^{(1)}(\omega)$ and $\Delta F_\lambda^{(1)}=(F_\lambda^{(0)})^2\phi_\lambda^{(1)}(\omega)$ are shown in Figs.~\ref{fig003a} and \ref{fig003b}, and the spin is conserved in each triangular vertex. Fig.~\ref{fig010} shows that the effective interaction $\Lambda_0^{a,b}$ between the hybridized particles occurs even in the leading contribution $\Delta\GD^{(1)}$ to $\GD^{(1)}$. In Fig.~\ref{fig010} $\Lambda_0^{a,b}$ are, however, dissolved in the bubble renormalizations of $B_\lambda^{(1)}$ and $F_\lambda^{(1)}$. This is unrelated to the $d$-$p$ anticommutation and we must therefore introduce the requirement that $\Delta\GD^{(1)}=\Delta\GD_M^{(1)}$.  In the SFT the spin factors of the first and the second term of Fig.~\ref{fig010}, associated with $\beta_\lambda^{(1)}(\omega)$ and $\phi_\lambda^{(1)}(\omega)$, are equal to 1 and 2 respectively. The time and the Pauli structure of the theory multiplies the first term by $-1$. The overall result is the subtraction $-1+2=1$ of the two relevant terms in Eq.~(\ref{Eq016}). In the MSFT the contribution of Fig.~\ref{fig010}a is removed by hand because it involves two bosons of equal spin, while the second term (Fig.~\ref{fig010}b) carries the spin weight 1 because the contribution with two bosons of equal spin is removed from the sum over the spins. This amounts to the $(F_\lambda^{(0)})^2\phi_\lambda^{(1)}/2$ renormalization of the $F_\lambda^{(0)}$ line. The overall result is again 1, i.e. $\Delta\GD^{(1)}=\Delta\GD_M^{(1)}$, provided that the prefactor of $\Lambda_0^b$ in Fig.~\ref{fig007}b, i.e. the singlet scattering in the MSFT $t_{pd}^4(\Lambda_0^a+\Lambda_0^b)$ is itself taken equal to the singlet scattering from the SFT. As required, the $d$-$p$ (anti)commutation rules are then entirely irrelevant in $\Delta\GD^{(1)}$ and $\Delta\GD_M^{(1)}$.

Next, the relation $\Delta\GD_M^{(1)}=\Delta\GD^{(1)}$ is to be extended to the Dyson theory which removes the double poles $(F_\lambda^{(0)})^2$ and $(B_\lambda^{(0)})^2$ from the above discussion. As previously discussed in the context of the SFT, $\GD^{(1)}$ is insensitive to the $p$-$d$ (anti)commutation rules. The relevant question is therefore whether or not the choice $\GD_M^{(1)}=\GD^{(1)}$, consistent with $\Delta\GD^{(1)}=\Delta\GD_M^{(1)}$, is unique, since the structure $B_\lambda^{(0)>}\ast (F_\lambda^{(0)})^2\phi_\lambda^{(1)}/2$  of $\Delta\GD_M^{(1)}$ may suggest otherwise, despite the fact that $\Delta\GD_M^{(1)}=\Delta\GD^{(1)}$.  However, expressing $\GD_M^{(1)}$ in terms of $F_\lambda^{(1)}$ and $B_\lambda^{(1)}$ ($f\leftrightarrow b$ symmetrically) allows the MSFT to benefit from the rapid convergence of the SFT towards the local gauge invariance and, in particular, to the LSR. In other words, by taking $\GD_M^{(1)}=\GD^{(1)}$ the singlet and triplet $\Lambda_0^{a,b}$ are taken again as dissolved, as in Fig.~\ref{fig010}, in the bubble renormalization of $B_\lambda^{(1)}$ and $F_\lambda^{(1)}$. At the $r=1,2$ stages of the Dyson perturbation MSFT it is thus irrelevant whether the double occupation of the Cu-site with equal spins is forbidden by the slave particles or by the Pauli principle. 
      
As already argued within the SFT, this line of reasoning can be extended within the NCA to $\GD^{(r-1)}=\GD_M^{(r-1)}$, i.e. to $\Gamma_{\vec k}^{(r)}$, because here the interaction vertices are also absorbed in Dyson bubble renormalizations of all the $f$-, $b$- and $p$-propagators involved. By definition, the NCA omits all except bubble renormalizations, and is therefore the same for the SFT and the MSFT. The first contribution in Fig.~\ref{fig008} thus obviously corresponds to the "exact" SFT/MSFT NCA for the $p$-, $f$-, and $b$-particles.
       
It is instructive to examine the modifications related to the lowest order SFT expression beyond the NCA, namely to $\GD^{(3)}$ discussed in connecton with Fig.~\ref{fig006} for $r=4$. The introduction of the $d$-$p$ anticommutation rules by the suppression of the triplet scattering of the physical fermions replaces ($\lambda$-independent) $\GD^{(3)}$ by ($\lambda$-independent) $\GD_M^{(3)}$. The difference $\GD^{(3)}-\GD_M^{(3)}$ is necessarily non-zero, because the interactions $t_{pd}^4(\Lambda_0^a+\Lambda_0^b)$ in Fig.~\ref{fig007} cannot be dissolved in bubble renormalizations of the slave particles. $\GD^{(3)}-\GD_M^{(3)}$ has a non-local and a local component. While the non-local component in question is related to the breakdown of the Cu-O anticommutation rules and should be discarded, the local component is involved in the SFT $n_{d\vec R+}=n_{d\vec R-}$ anticommutation rule, as understood in Eq.~(\ref{Eq007}). The latter is thus affected in the MSFT beyond the $r=4$ level of iteration even though the Dyson structure of Eq.~(\ref{Eq005a}) is maintained. On the other hand, the MSFT is taken to keep the empty d$^8$ state of Eq.~(\ref{Eq007}) at "infinite" energy $\varepsilon_d+U_d$ as available for the creation of an additional fermion on the Cu site. This can be extended to $r\geq4$ in Eqs.~(\ref{Eq005a}) and (\ref{Eq005b}) for $\GD_M^{(r)}$ and $\Gamma_M^{(r)}$, obtained by replacing $\GD^{(r-1)}$ by $\GD_M^{(r-1)}$ and omitting the contributions of the triplet scatterings. In order to satisfy the equality $n_d^{(r)-}=n_d^{(r)+}$ for $r\geq4$ one should thus allow for the additional spectral density at the energy $\varepsilon_d+U_d$ on the $t<0$ side. This means that in Eq.~(\ref{Eq007}) $\sigma_M^{(r-1)>}$ differs from $\sigma^{(r-1)>}$ of the SFT for $r\geq4$.

In other words, it can be reasonably expected that the MSFT is locally gauge invariant on average and, to a good approximation, $Q_{MR}^{(r)}\approx1$ if the $f\leftrightarrow b$ symmetry is maintained in the calculation of the $f$- and $b$- propagators. One can even expect that the MSFT is a better "conserving" approximation\cite{kr1} than the NCA because it has the advantage that it generates {\it incommensurate} magnetic correlations even for finite $t_{pp}$. The asymptotic local gauge invariance is however conserved only if the omission of the triplet scatterings between any two $pdp$-particles, corrected for the lack of the local spectral weight by Eq.~(\ref{Eq007}), renders the ground state exactly antisymmetric. The rigorous proof of this assertion is lacking so far. Nevertheless, the MSFT has some advantages over the DMFT based on small clusters\cite{zl1,mc1} of CuO$_2$ units, because this latter is appropriate only for the description of commensurate structures. Unfortunately, the DMFT approach which allows perturbatively for incommensurate magnetic correlations\cite{rb1} is at present restricted to the single band Hubbard model.

\section{Ladder approximation for coherent magnetic correlations\label{Sec10}}

The present theory based on the $U_d\rightarrow\infty$ Hamiltonian of Eq.~(\ref{Eq003}) generates the d$^{10}$$\leftrightarrow$d$^9$ disorder in the $r=2$ single particle propagators and the singlet $pdp$ particle-particle and particle-hole correlations for $r=4$, as shown in Fig. \ref{fig006}. Since the effective interaction $t_{pd}^4\Lambda_0$ turns out to be repulsive for long times its primary effect concerns the spin flip electron-hole correlations in addition to the d$^{10}$$\leftrightarrow$d$^9$ disorder. Although the d$^{10}$$\leftrightarrow$d$^9$ disorder occurs in the single particle propagation in lower order than the spin flip processes, the leading $r=2$ disorder contribution falls far from the Fermi level and the subsequent $r=3$ disorder term expands in terms of $n_d$, assumed small according to Eq.~(\ref{Eq015}). It should be noted in this respect that the choice of the iterative renormalization scheme is not unique, i.e. that other appropriate prescriptions can be devised according to the value of $x$ and the band parameters under consideration. E.g., the NCA emphasizes the d$^{10}$$\leftrightarrow$d$^9$ charge-transfer disorder effects and neglects the magnetic correlations. In contrast, the choice of the free $F^{(0)}$ and $B^{(0)}$ in Fig.~\ref{fig006}b together with the hybridized $pdp$ propagators $t_{pd}^{-2}\Gamma_{\vec k}^{(1)}$ obviously favors the strong coherent SDW correlations for $x\approx x_{vH}$. Notably, this flexibility in the selection of the perturbation subseries has its physical counterpart because magnetic coherence is expected to compete with the d$^{10}$$\leftrightarrow$d$^9$ disorder, and win when it is sufficiently strong. This suggests the investigation of Fig.~\ref{fig006}b using the free $F^{(0)}$, $B^{(0)}$ and $t_{pd}^{-2}\Gamma_{\vec k}^{(1)}$ propagators.

In this context, it is interesting to examine the $pdp$-$pdp$  magnetic spin-flip susceptibility $\tilde\chi_{SDW}$ in the coherent limit. The first relevant spin-flip correlation is associated with the free $pdp$-$pdp$ bubble $\chi_{pp}^{(1)}(\vec q,\omega)$. For $\omega$ small it is dominated by the intraband contribution $\chi_{pp}^{LL}(\vec q,\omega)$. The latter should be distinguished from the $dpd$-$dpd$ bubble with the intraband contribution $\chi_{dd}^{LL}(\vec q,\omega)$. Both functions exhibit\cite{bb1,sk3} the same {\it singular} behavior in the reciprocal space when $\mu^{(1)}$ falls close enough to the band van Hove singularity but differ in the associated spectral densities.

\begin{figure}[tb]

\begin{center}{\scalebox{0.5}
{\includegraphics{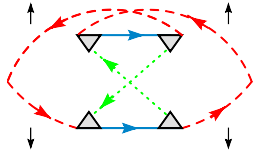}}}
\end{center}

\caption{(Color online) $t_{pd}^4\Lambda_0^a$ renormalization of the magnetic spin flip susceptibility; spin (vertical arrows) is conserved in triangular vertices.\label{fig011}}

\end{figure}

$\chi_{pp}^{LL}$ is further renormalized, in particular according to Fig.~\ref{fig011} where $t_{pd}^4\Lambda_0^a$ appears explicitly as the effective singlet interaction between the $r=1$ $pdp$ propagators.\cite{bb1} An analogous contribution comes from $t^4_{pd}\Lambda_0^b$. The actual calculations are simplified by the separability of the interactions $t_{pd}^4\Lambda_0^{a,b}$, like in Eq.~(\ref{Eq014}). The eight poles involved are split in $4+4$ sets. The real part of the integrals is approximately proportional to $(\chi_{pp}^{LL})^2t_{pd}^4\Lambda_0$ in agreement with the estimate of Eq.~(\ref{Eq015}) (note that this results in cancellations of numerous $\Delta_{d\mu}$ factors between the squared spectral weights $z_{\vec k}^{(L)}$ of Eq.~(\ref{Eq008}) and $t_{pd}^4/\Delta_{d\mu}^3$). Apparently, the retarded nature of $t_{pd}^4\Lambda_0$ does not play an essential role in this result. $(\chi_{pp}^{LL})^2t_{pd}^4\Lambda_0$ thus increases the bare $\chi_{pp}^{LL}$. This latter can be large at finite $x$ due to nesting\cite{bb1,sk3} and the ladder approximation then leads to strong low energy SDW correlations $\tilde\chi_{SDW}$,

\begin{equation}
\tilde\chi_{SDW}\approx\frac{\chi_{pp}^{(1)}}{1-U_{d\mu}\chi_{pp}^{(1)}}\label{Eq017}\;,
\end{equation}

\noindent where $U_{d\mu}\approx4t_{pd}^4/\Delta_{d\mu}^3$ of Eq~(\ref{Eq015}). Equation (\ref{Eq017}) corresponds to the simplest vertex renormalization (which omits to single out the Umklapp component\cite{dz2} of $U_{d\mu})$ in the symbolic equality shown in Fig.~\ref{fig008}. 

Such a distribution of the roles of the propagators and interactions contrasts with the small $U_d$ theory for $\tilde\chi_{SDW}$, where the interaction $U_d$ involves the $dpd$ bubbles $\chi_{dd}^{LL}$. So does the ansatz\cite{yi1,km1} $U_d\rightarrow J_{pd}(\cos k_x+\cos k_y)$ with $J_{pd}=4t^4_{pd}/\Delta_{pd}^3$ that is traditionally used in the approach from the MFSB side. Note the essential role of the negative sign associated with the $\cos$ term for $\vec k=[\pi,\pi]/a$ and questioned in Ref.~\onlinecite{an2}. Equation~(\ref{Eq017}) resolves thus the long lasting\cite{yi1,an2,km1} controversy of what replaces $J_{pd}=4t_{pd}^4/\Delta_{pd}^3$ of the $U_d=\infty$ Mott-AF limit in the $U_d=\infty$ metallic limit. The answer is that the quantity in question is {\it the repulsion} $U_{d\mu}\approx4t_{pd}^4/\Delta_{d\mu}^3$ {\it between the $pdp$ propagators}.

In cuprates this replacement occurs progressively by doping through $x_{cs}$ as discussed in Sec.\ref{Sec08}. All contributions to the magnetic susceptibility $\chi_{SDW}$, those stemming from $\chi_{bb}^{\uparrow\downarrow}$ of Eq.~(\ref{Eq022zvj}) and the cross terms should in principle be treated on equal footing with $\tilde\chi_{SDW}$ near above $x_{cs}$. However, $\tilde\chi_{SDW}$ dominates the coherent fluctuations. As long as $|\mu-\omega_{vH}|$ is large, so that the "nesting" effects in $\chi_{pp}^{(1)}(\vec q,\omega)$ are weak, the spectral weights $z_{\vec k}^{(L)}$ of Eq.~(\ref{Eq008}) may be important in this respect. Under the additional assumption that $|\mu-\omega_{vH}|$ is small $\tilde\chi_{SDW}$ of Eq.~(\ref{Eq017}) becomes large at $\omega$ small and its singular behavior determines then alone the low frequency behavior of the system. 
      
A similar line of reasoning, invoking the effective, instantaneous repulsion between (simply renormalized) single particle propagators, with additional corrections, can be applied to the discussion of many observed properties of cuprates.\cite{fr1,bs5,dz2,fr3} It is therefore of some interest to discuss briefly how the present theory deals with some salient experimental results. 

\section{Comparison with experiments}

\subsection{ARPES and NQR\label{Sec11.1}}

ARPES experiments\cite{in1,yo1,vl1,zu1}  measure the dispersion properties of the exact electron propagators in the reciprocal space which are related by the sum rule to $n_d$ and $n_p$ determined from the NQR data.\cite{ku1} They show unambiguously that for $x>x_{cs}$ $\mu$ falls in the {\it broad} band, below the Cu site energy $\tilde\varepsilon_d$. As in the MFSB case,\cite{mr1} the nonmagnetic (unreconstructed\cite{se1}) band structures with "Fermi arcs on large surface" are  observed for $x>x_{cs}$. They are first fitted here to determine two renormalized band parameters, conveniently $\tilde t_{pd}/t_{pp}$ and $\tilde\Delta_{pd}/t_{pp}$, while $t_{pp}$ is chosen finally to fix the overall energy scale. The observed Fermi energy $\mu$ is conveniently measured by $\omega_{vH}-\mu$ where $\omega_{vH}$ is the position of the vH singularity. In parallel, the NQR measurements determine the absolute values of $n_d(x)$ with a few percent accuracy due to the ambiguity in the evaluation\cite{ku1} of the Madelung contribution to the electric field gradients.
 
The measured band parameters are identified here with those calculated in Eqs.~(\ref{Eq012b}), (\ref{Eq012a}) and (\ref{Eq015zvj}) for $r=2$, namely with $t_{pd}^{(1)}/t_{pp}$ and $\Delta_{pd}^{(1)}/t_{pp}$ (keeping in mind that $\varepsilon_p$ and $t_{pp}$ are not renormalized). For comparison with experiments the latter should include iteratively or selfconsistently,\cite{tu1} primarily through the site energies $\varepsilon_d^{(1)}$ and $\varepsilon_p$, the variation of the Madelung potential due to the variation from $n_d^{(1)}$ to $n_d^{(2)}$. Those parameters, together with $\omega_{vH}^{(1)}-\mu^{(2)}$, determine $n_d^{(2H)}$ and $x_{eff}$ of Eq.~(\ref{Eq013b}). In principle the bare parameters $t_{pd}/t_{pp}$, $\Delta_{pd}/t_{pp}$ and $\mu^{(1)}$ can be found from $t_{pd}^{(1)}/t_{pp}$, $\Delta_{pd}^{(1)}/t_{pp}$ and $x_{eff}$ under assumption that renormalizations are weak. In practice this procedure implies a heavy numerical work required by the inversion of Eqs.~(\ref{Eq012a}) and (\ref{Eq012b}). Thus, instead of calculating $n_d^{(2)}$ from ARPES the latter is estimated from NQR measurements of $n_d$ and $n_p$. In this spirit we discuss below two different $r = 2$ regimes, associated respectively with lanthanates and YBCO. 
      
The example of an excellent, essentially two-parameter fit obtained from the three band model is given in Fig.~\ref{fig004} for $x=0.07$ LSCO. Typical values of the band parameters satisfy $\tilde\Delta_{pd}^2>2\tilde t_{pd}^2\approx\tilde\Delta_{pd}|t_{pp}|>4t_{pp}^2$, the regime with relatively flat Fermi arcs. The observed $\mu$ corresponds to $n_d^{(2H)}\approx1/2$. Upon doping $x=x_{vH}\approx1/5$ ($1/8$ in\cite{vl1} LBCO)  $\mu$ traverses\cite{in1,yo1} the vH singularity. The Luttinger sum rule is qualitatively satisfied\cite{yo1} indicating that $n_d^{(1)2}/2\approx n_d^{(2inc)}$ cancellations in Eq.~(\ref{Eq013b}) lead to $x_{eff}\sim x$. Indeed, good band fits can then be obtained in function of $x$, as already noted before\cite{mr1} in the MFSBT. The renormalizations $t_{pd}^{(1)}/t_{pp}$ and $\Delta_{pd}^{(1)}/t_{pp}$ of band parameters by $n_d^{(1)}$ are significiant, but NQR gives\cite{ku1} $n_d\leq3/4$ and a {\it positive} $\partial n_d/\partial x\approx1/4$. Although consistent with the decrease of $t_{pd}^{(1)}/|t_{pp}|$ and increase of $\Delta_{pd}^{(1)}/|t_{pp}|$ upon doping, we feel that the resulting difference between $n_d\approx n_d^{(2)}$ and $n_d^{(2H)}$ somewhat stretches the low order theory.
      
For YBCO$_{6+\delta}$ NQR gives\cite{ku1} $n_d\approx1/2$, $\partial n_d/\partial\delta\approx1/10$ and $x\approx\delta/5$ for $x>x_{cs}\approx0.05$, while the observed\cite{sch} Fermi level $\mu$ falls well below the vH singularities (of the band doublet) at $\tilde\omega_{vH}-\mu\approx0.25$ eV which does not depend much\cite{rl1} on $\delta$. The observed $\tilde\Delta_{pd}/|t_{pp}|$ and $\tilde t_{pd}/|t_{pp}|$ are smaller than in lanthanates, towards the regime $\tilde\Delta_{pd}^2\approx2\tilde t_{pd}^2\approx\tilde\Delta_{pd}|t_{pp}|\approx4t_{pp}^2$ exhibiting pronounced arcs of the unreconstructed\cite{se1} Fermi surface. Consequently, the $r =2$ renormalizations of band parameters are weaker and $n_d^{(1)}$ is smaller than in lanthanates, consistently with large $\omega^{(1)}_{vH}-\mu^{(2)}$. The negative incoherent corrections to $x_{eff}$ of Eq.~(\ref{Eq013b}) tend then to be large compared to $n_d^{(1)2}/2$. Notably, $x_{eff}\approx x-n_d^{(2inc)}\approx0$ means that doped holes go at average mostly to the highly degenerate set of disordered d$^9$ states,\cite{go3} which is consistent with $\partial(\omega_{vH}^{(1)}-\mu^{(2)})/\partial x\approx0$. The ARPES result\cite{rl1} $\partial(\tilde\omega_{vH}-\mu)/\partial\delta\approx0$, traditionally\cite{sch} named breakdown of the Luttinger sum rule (assuming a reasonable $x\sim\delta$), can thus be brought into broad agreement with the theory.
      
While ARPES provides $d(\mu-\omega_{vH})/d\delta$ for $x>x_{cs}$, it is generally believed\cite{mi1} that XPS in cuprates gives the absolute shift $d\mu/d\delta$ itself. It turns out\cite{mi1} so that in metallic YBCO$_{6+\delta}$ ($\delta\approx5x$), $d\mu/d\delta<0$ (in the hole language) and tends to vanish when optimal doping $\delta\approx1$ is approached. On the other hand, for metallic LSCO, the shift $d\mu/d\delta$ ($\delta\approx x$) is found\cite{mi1} small all along. This issue is discussed here by setting conveniently
      
\begin{equation}
\mu^{(2)}=\mu^{(2)}-\omega_{vH}^{(1)}+\bar\varepsilon^{(1)}-\frac{1}{2}\sqrt{\Delta_{pd}^{(1)2}+16t_{pd}^{(1)2}}\label{Eq028zvj}
\end{equation}
      
\noindent where $\bar\varepsilon^{(1)}$ is the average site energy $\bar\varepsilon^{(1)}=(\varepsilon_d^{(1)}+\varepsilon_p)/2$. As easily seen such expression cannot reconcile the ARPES data and the XPS analysis unless the variations with $x$ of the in plane Madelung field produced by the negative charge $x$ of dopants are taken into account. In particular, the component of the Madelung field which is uniform within, and among, the CuO$_2$ unit cells produces the variation of $\bar\varepsilon^{(1)}$ with $x$, while $\Delta_{pd}^{(1)}$ ($\Delta_{pp}=\varepsilon_{px}-\varepsilon_{py}=0$) are less affected. The estimation of the average Madelung field is simple when the negative charge is distributed in homogeneous layers of surface density $x/a^2$. The negative shift $\delta^M \bar\epsilon$ of $\bar\varepsilon^{(1)}$ (conjugated\cite{bs2} to $n_d+2n_p$) of the order od $xV_{pd}c/a$ for lanthanates, unscreened within the plane, tends to cancel out the positive variation of $\mu$ due to the filling of the CuO$_2$ plane with $x$ holes at fixed $\omega_{vH}^{(1)}$. This agrees broadly with the XPS analysis.\cite{mi1} 

Notably further, $\partial\mu/\partial x\propto(\delta(n_d+2n_p))^{-2}\ll1$ indicates\cite{tu1} a propensity of the normal metallic phase to inhomogeneous phase separation in the plane. Macroscopic charge redistribution is however hindered\cite{bs2,tu1,ku3} by long range Coulomb screening. It is thus appropriate to consider the charge transfers over short distances, among or within the CuO$_2$ unit cells, as will be discussed next.

\subsection{Magnetic orderings and lattice deformation structures}

As mentioned in the Introduction, various experimental methods,\cite{ts1,gz1} including NMR,\cite{al1,go3} STM,\cite{sg1} X-ray\cite{bi1} and quasi-elastic neutron scattering\cite{tr2,st1} in particular, reveal entangled,\cite{bi1,sg1} (quasi-)static magnetic and lattice deformation structures, which break, or restore as a checkerboard, the D$_4$ symmetry of the CuO$_2$ lattice. They are characterized by the Bragg spots which occur at $\vec q_{SDW}$ and $\vec q_0$ respectively and, assuming the satellites to those {\it leading} harmonics, are usually called stripes.\cite{ki1,vt1} At small $0<x<x_{cr}\approx0.05$ $\vec q_0$ and $\vec q_{SDW}$ lie on the zone diagonals but for doping $x>x_{cr}$ they are rotated by $\pi/4$ and become collinear with the main zone axes.\cite{ts1} The relation 
      
\begin{equation}
\vec q_0+2\vec q_{SDW}=\vec G\label{EqS}
\end{equation}
 
\noindent is satisfied all along.\cite{gz1,ki1,vt1} In addition to stripes, LBCO-like cuprates exhibit\cite{ax1,kmr,tr3} the {\it commensurate} LTO/LTT lattice phase transition.

The present approach attributes diagonal stripes to the weakly doped Mott-AF phase. Indeed, it is likely that a coherent $t,t',J,J'$ model\cite{cl1} can roughly\cite{wl1} explain such stripes, though we are not aware of any previous to ours attempt\cite{sk3} to relate $t', J'$ etc. to parameters of the Emery model. On the other hand the present theory of the metallic phase should produce the collinear stripes or checkerboard configurations which satisfy $\vec q_0^{coll}+2\vec q^{coll}_{SDW}=\vec G$.
      
\begin{figure}[htb]

\begin{center}
{\scalebox{0.31}{\includegraphics{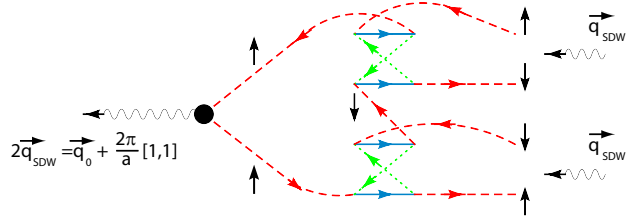}}}
\end{center}

\caption{(Color online) Umklapp contribution due to two SDW's at $\vec q_{SDW}\approx\vec G/2$, resulting in momentum transfer $\vec q_0$. The vertex denoted by the dot depends on the symmetry of the intracell fluctuation or phonon which carries the momentum $\vec q_0$ away.\label{figA}}

\end{figure}
      
When dealing with this issue we realize first that external frequencies in the involved correlation functions can be taken to vanish (adiabatic approximation), i.e. it is then possible to construct the corresponding Landau-like theory. In contrast to $\chi_{bb}^{(1)}(\vec q,0)$ of Eq.~(\ref{Eq022zvj}) $\chi_{pp}^{(1)}(\vec q_{SDW}\approx\vec G/2,0)$ of Eq.~(\ref{Eq017}) exhibits a log-like singularity at $\omega\approx0$ which is the strongest when $|\omega_{vH}-\mu|$ is small enough, as it certainly is in lanthanates, as we choose from now on for brevity.  Importantly, $\chi_{pp}^{(1)}(\vec q,\omega)$ may then present\cite{qs1,yi1,sk3} peaks for incommensurate $\vec q_{SDW}^{coll}$ in the vicinity of $\vec G/2$ even when $t_{pp}$ is finite. This can be taken to generate the Landau theory in which the $\vec q_{SDW}^{coll}$ SDW is the primary order parameter. The lattice deformations are driven by two such SDW's through the Landau invariant shown in Fig.\ref{figA} obtained on assuming that all external frequencies are small. It can be easily seen by direct inspection of Fig.~\ref{figA} that when $|\omega_{vH}-\mu|$ is small the main contribution comes when three $pdp$ propagators which form the triangle fall into the vicinity of the vH points. Insertion of the B$_{1g}$-like vertex on the top of the triangle in Fig.~\ref{figA} is then unessential, i.e. the two $\vec q_{SDW}^{coll}$-SDW's produce a large B$_{1g}$-like O$_x$/O$_y$ CT with the wave vector $\vec q_0^{coll}$ which satisfies Eq.~(\ref{EqS}), while the B$_{2g}$-like CT is small. The induced A$_{1g}$-like CDW on the total charge of the CuO$_2$ unit cell is also small for $\vec q_0$ small\cite{bs2,ku3} and moreover screened\cite{ku3} by long range Coulomb forces. Unscreened, the latter is sizeable for arbitrary $\vec q_0$ and, in particular, interacts\cite{bs2} by the e-p interactions with LTT modes. Moreover, the LTT tilts are entangled, either via O$_x$/O$_y$ CT or directly by ionic interactions,\cite{pu1} with the $\vec q_0^{coll}$ shears of the CuO$_2$ plane. A sizeable $\vec q_0^{coll}$ is necessarily associated with {\it linear} coupling of the lattice displacements, such as is the oxygen LO phonon,\cite{pnc} to the mixture of O$_x$/O$_y$, Cu/O$_2$  CT's  and CDW (including the bond density waves not discerned here). Let us note on the semi-quantitative side that even when $\vec q_0^{coll}$ is close to the CuO$_2$-lattice tetramerization $2\pi[1/4,0]/a$, the omission of Coulomb screening in the straightforward MFSB calculation\cite{lo2} of stripes overestimates the $\vec q_0^{coll}$-CDW with respect to the $\vec q_0^{coll}$-O$_x$/O$_y$ and Cu/O$_2$ CT's.
      
The observed $\vec q_0^{coll}$ is large with respect to the inverse SDW correlation length,\cite{st1,tr2,gz1} i.e., the formation of stripes does not interfere to the lowest order with the behavior of $\chi_{pp}^{(1)}(\vec q,\omega)$ at $\vec q\approx0$. $\RE\chi_{pp}^{(L,L)}(\vec q,0)$ is logarithmically large at $\mu\approx\omega_{vH}$ ($x\approx x_{vH}$) in lanthanates\cite{in1,yo1} not only at $\vec q_{SDW}$ but also at $\vec q=0$. This corresponds to the Jahn-Teller (JT) splitting of the vH singularities\cite{fr1,bs2,ku3} rather then to their nesting, relevant for the SDW case with $|\omega_{vH}-\mu|$ small. Such JT splitting produces the homogeneous $\vec q\approx0$ intracell B$_{1g}$-O$_x$/O$_y$ CT $n_{px}-n_{py}$, while the $B_{2g}$-contributions and the overall A$_{1g}$-CDW are {\it absent} by symmetry.\cite{bs2,ku3} Clearly,\cite{ku3} the dynamic B$_{1g}$-O$_x$/O$_y$ CT is accompanied by local though not circular\cite{va2} currents. $\chi_{pp}^{(1)}(\vec q=0,\omega)$ (without spin flip) has no direct vertex renormalizations by large $U_d$ (in particular the Coulomb interaction $V_{pp}$ between two neighboring oxygens, may be invoked for further enhancement of $\chi_{pp}^{(1)}$). The $\vec q=0$ O$_x$/O$_y$ CT is coupled by the {\it quadratic} e-p coupling to two LTT-tilts with $\vec Q=2\pi[1/2,0]/a$ each, since the linear e-p coupling then vanishes\cite{bs2} by symmetry. Such LTT dimerization (which produces the splitting $\Delta_{pp}=\varepsilon_{px}-\varepsilon_{py}$ of  the oxygen site energies, conjugated\cite{bs2} to $n_{px}-n_{py}$)  plays the role of the primary order parameter\cite{bs2} of the first order LTO-LTT phase transition. The latter gives rise in particular to the {\it new} Bragg spot at $2\vec Q=2\pi[1,0]/a$, as it should,\cite{ft1,tr3} accompanied by the stripe leading harmonic at $\vec q^{coll}_0$.
           
The above description agrees remarkably well with the observed\cite{bi1,sg1} SDW-LTO-LTT structure of the striped phase in lanthanates, including the first order LTO-LTT phase transition in LBCO.  It emphasizes the importance of sizeable O$_x$/O$_y$ CT $n_{px}-n_{py}$ between two oxygens and is thus consistent with appreciable charge sharing between Cu and 2O invoked here. In contrast, it is obvious that those results and in particular the simple O$_x$/O$_y$ JT effect, can hardly be explained by the rigid $t-J$ model which keeps the average (oxygen) occupation of the {\it upper} band low, proportional to $x$.  

\subsection{Inelastic spin-polarized neutron scattering}

The inelastic spin polarized scattering of {\it cold} to {\it thermal} neutrons is usually taken to measure the exact $\chi^{\uparrow\downarrow}(\vec q,\omega)$, associated with the flip of the {\it overall} spin in the CuO$_2$ unit cells. This fixes the relative phases of $\chi^{\uparrow\downarrow}_{dd}$, $\chi^{\uparrow\downarrow}_{pd}$ and $\chi_{SDW}$ in $\chi^{\uparrow\downarrow}(\vec q,\omega)$. The large scattering for $\omega$ around and above 0.15 eV occurs\cite{st1,tr2}  for $\omega(\vec q)$ which differs clearly\cite{tr2} from the AF-magnon parent dispersion\cite{cl1} with $J_{pd}\approx0.15$ eV. In addition the dispersion is considerably broadened and presents the absolute maximum of intensity for $\vec q_{SDW}^{dg}(\omega)$ on zone diagonals. As concerns $\tilde\chi_{SDW}(\vec q,\omega)$ of Eq.~(\ref{Eq017}), its properties result from the interplay in $\chi_{pp}^{(1)}(\vec q,\omega)$ of "small" energy scales, $\omega$, $\omega_{vH}-\mu$, $t_{pp}$ and $t_{pd}(\vec q-\vec G/2)$. A systematic analysis of $\tilde\chi_{SDW}(\vec q,\omega)$ is currently under way,\cite{sk3} but early numerical calculations\cite{xu1} of $\chi^{(1)}(\vec q,T)$ for reasonable values of band parameters indicate that $\tilde\chi_{SDW}(\vec q,\omega)$ may be enhanced at appropriate values of $\vec q_{SDW}^{dg}(\omega)$. The corresponding energy transfers $\omega$ are also possibly affected by $\chi^{\uparrow\downarrow}_{bb}$ associated with Eq.~(\ref{Eq022zvj}) and its RVB extensions. Further on, serious departures from the Heisenberg AF-behavior occur at lower frequencies, especially below $\omega_{hg}\approx50$ meV where the so called hourglass dispersion is observed.\cite{tr2} This indicates that $\chi_{SDW}(\vec q,\omega)$, which includes the single particle (pseudo)gapping,\cite{yi1} due here to the SDW itself, is important in the frequency range $\omega_{hg}\geq\omega>\omega_{ph}$ where $\omega_{ph}$ are the bare frequencies of phonons involved in static stripes. In this range $\chi_{SDW}(\vec q,\omega)$ decouples from the lattice (but not from the O$_x$/O$_y$ CT in Fig.~\ref{figA}) and an interesting open question is whether it by itself tends to break or not the D$_4$ symmetry of its ladder approximant $\tilde\chi_{SDW}(\vec q,\omega)$. The previous analytic\cite{sc1} and numeric\cite{qs1,yi1} calculations for such $\omega$ show that $\tilde\chi_{SDW}(\vec q,\omega)$ presents peaks at $\vec q_{SDW}^{coll}$ on the main axes, in agreement with the observed $\pi/4$ rotation of the absolute maxima in the neutron scattering. Finally, for $\omega<\omega_{ph}\approx12$ meV one observes\cite{tr2,si3} the conic dispersion of spin-lattice density waves, expected in generalization of the present approach which takes the SDW at $\vec q_{SDW}^{coll}(\omega\approx0)$ as the primary order parameter of the stripe formation. 

\subsection{Magnetic pseudogap}

Beyond the unreconstructed Fermi surface, such as that in Fig.~\ref{fig004}, the ARPES measurements\cite{lu1,zu1,vl1} reveal in lanthanates the pseudogap around the antinodal (vH) points of the Brillouin zone, while in YBCO-like (and electron doped) materials the strong spectral density variations\cite{kk1} are observed along the unreconstructed Fermi surface, together with shadow bands.\cite{se1,sch} These structures are usually associated with the magnetic energy scale of about 50 meV, reminiscent of $\omega_{hg}$ associated above with the hourglass dispersion. Such ARPES structures were discussed before\cite{ru1,no1,sk2} in some detail, assuming that the propagating  hole emits and annihilates a damped soft AF-paramagnon described by a phenomenological $\chi^{\uparrow\downarrow}(\vec q\approx\vec G/2,\omega)$ which couples to the propagating hole by an adjustable coupling constant. Although incommensurability and couplings to the lattice are omitted in $\chi^{\uparrow\downarrow}$, those approaches reproduce observations remarkably well.\cite{ru1,sk2,no1} In particular the adjustable coupling constant of those calculations is associated here by Fig.~\ref{fig006} and Eq.~(\ref{Eq008}), with the microscopically derived $U_{d\mu}$ of Eq.~(\ref{Eq017}) involved also in $\tilde\chi_{SDW}$. This lends a semi-qualitative support to "paramagnon" approaches\cite{ru1,sk2,no1} to the magnetic pseudo gap.

\subsection{Raman, optic and soft-X-ray measurements}

The compelling evidence for the crossover picture proposed here comes from the Raman and optic experiments. We notice in this respect on the expected {\it magnetic} scale up to 0.5 eV that the energy of the sharp two-magnon resonance observed\cite{si2} at $x=0$ in the B$_{1g}$ Raman data for lanthanates is about 0.35 eV. This Raman activity arises from the coupling\cite{lo1} of the two-magnon resonance of the Heisenberg model on Cu sites with super exchange $J_{pd}$ (via O-sites) to the $\vec q=0$ B$_{1g}$ quadrupolar O$_x$/O$_y$  CT within the CuO$_2$ unit cell (and further to photons). Measurements show clearly\cite{bl1,si2} that the two-magnon resonance shifts downwards and fades progressively away for $x>x_{cs}\approx0.05$. This is necessarily related to the energy shift, broadening and loss of the spectral density of magnons involved in the two-magnon resonance. The continuity through $x_{cs}$ suggests that the broad structure observed for $x>x_{cs}$ reflects the short range AF-RVB fluctuations generated by Eq.~(\ref{Eq022zvj}). Indeed, the contribution of two SDW's each described by $\tilde\chi_{SDW}(\vec q,\omega)$ coupled according to Fig.~\ref{figA} to the quadrupolar $\vec q=\vec q_0$ O$_x$/O$_y$ charge CT within the CuO$_2$ unit cell is not expected to produce large Raman response at the energies as high as $0.35$~eV. Actually the $pdp$ triangle in Fig.~\ref{figA} is dominated for $|\omega_{vH}-\mu|$ small by three propagations in the vicinity of the vH points, i.e. this contribution is expected to be large only for lower frequencies, and then at finite $\vec q_0(\omega)\neq0$. Rather than with Fig.~\ref{figA} the Raman $\vec q=0$ B$_{1g}$ response at frequencies below $\omega_{hg}\approx50$ meV is thus dominated by the $\vec q=0$ O$_x$/O$_y$ CT associated to the Jahn-Teller $\chi_{pp}^{(1)}(0,\omega)$ with pseudogap corrections. So far however there is no direct evidence for the coherent quadrupolar exciton (in the whole Raman range of frequencies) and the O$_x$/O$_y$ CT fluctuations are observed through their coupling to phonons. In YBCO this concerns\cite{si2,ts2} e.g. the $\vec q=0$ B$_{1g}$ phonon at 340 cm$^{-1}$ and in lanthanates the $\vec Q$ LTT-phonons, coupled in pairs to Raman photons via the intracell quadrupolar O$_x$/O$_y$ CT-fluctuations. This assignation agrees qualitatively with sharp and two phonon-broadened features observed respectively in $x\approx0.1$ YBCO and lanthanates by low frequency Raman\cite{ts2} scattering.
      	
Similarly, the optic conductivity of conducting lanthanates exhibits a hump\cite{uc1} at $\omega\approx0.5$ eV for $0<x<0.05$ which shifts downwards, broadens and eventually fades away upon doping through $x_{cs}\approx0.05$. This hump may well correspond to the transitions of $x$ holes involved in the Zhang-Rice singlets between the bonding and nonbonding states on oxygens\cite{za1} It is reasonable to think for $x>x_{cs}$ that the (downshifted and broadened) 0.5 eV optical hump\cite{uc1} is the RVB remnant of the Mott-AF phase, in parallel with that observed\cite{si2} around $0.35$~eV in the B$_{1g}$ Raman scattering, while the rest can be explained\cite{ku2,ku3} by the interband optic transitions, such as those of Fig.~\ref{fig004}, of the weakly renormalized three-band model. Indeed, no resonant optical structure related to the 3-band model appears in the $0.5$~eV - $50$~meV range of our $r=2$ calculation. Notably, the crossover from the Mott-AF state to the metallic state is accompanied in such approach by the shift of chemical potential with $x$ from that associated with small $x$ Zhang-Rice structure to that appropriate for the moderately renormalized 3-band model. As concerns the conductivity at frequencies below $\omega_{hg}\approx50$~meV, it is described here basically in terms of (no spin flip) $\chi_{pp}^{(1)}(0,\omega)$ with nonmagnetic vertex corrections, including\cite{ku2,ku3} long range Coulomb screening. This picture may also take into account the self-energy renormalizations of the $pdp$-propagators associated with magnetic (pseudo\cite{ru1,no1,sk2}) gap\cite{ku3} scale and perturbative vertex corrections associated to {\it nonmagnetic} Umklapp scatterings, such as that arising from the first and second neighbor Cu-O interactions $V_{pd}$ and $V_{pp}$. The vH antinodes are however unimportant for low frequency conductivity, because the carrier mobility vanishes there with or without a pseudogap. In contrast to that the vicinity of nodal points is not\cite{sk2} affected by the pseudogap at least for sizeable $t_{pp}$, giving rise to a conduction picture, which, in the simplest possible low T approximation, is associated with the Fermi liquid Umklapp T$^2$ law in the dc resistivity. Interestingly, by using sum rules appropriate for coherent electron-hole excitations,\cite{ku3}  the described picture for infrared conductivity and Raman activity can be brought in good agreement with the Hall effect measurements\cite{uc1,on1,go3} in weak magnetic fields, while the corresponding quantum oscillations\cite{se1} are currently under consideration from the present point of view.

The soft-X-ray absorption is associated with the creation of electron-hole pairs, with the hole created in the deep core state. Actually, an early observation\cite{ch1} of the crossover at small  $x_{cs}$ was accomplished using this method. The interpretation\cite{ch1} of the data was carried out using the Zhang-Rice $t-J$ limit\cite{za1} for {\it all} $x$ of interest.  For $x=0$ and $x>0$ holes are annihilated respectively in the {\it lower} level and in the {\it upper} level (narrow ZR band), using the hole language. While we keep the same picture for $x\leq x_{cs}$, we associate here the X-ray absorption in the $x\geq x_{cs}$ metallic phase with the transition of the hole from the renormalized {\it lower} $L$-band to the core level.  The difference $\Delta E$ between the (low order) hole energies for the $x\leq x_{cs}$ and $x\geq x_{cs}$ regimes is,\cite{bbk,mr1,sk4} in particular, according to Eq.~(\ref{Eq028zvj}),

\begin{equation}
\Delta E =\bar\varepsilon-\bar\varepsilon^{(1)}+\frac{1}{2}\left(\sqrt{\Delta^{(1)2}_{pd}+16t_{pd}^{(1)2}}-\sqrt{\Delta_{pd}^2+16t_{pd}^2}\right)\;,
\end{equation}

\noindent taking for simplicity that $x_{cs}=0^+$ and $t_{pp}=0$. This takes into account that, unlike in the metallic phase, the intercell hybridization is forbidden in the $x=0$ N\' eel phase. Appreciable renormalization of the Cu-O hopping in Eq.~(\ref{Eq015zvj}), rather than of the average site energy $\bar\varepsilon$ and the CT gap $\Delta_{pd}$, consistent with Fig.~\ref{fig004}, makes $\delta E$ negative. The latter is also approximately equal to the negative shift in the luminescence frequency, as originally observed.\cite{ch1} Although such a trend is satisfactory it is appropriate to mention that, when the variations of $n_d-2n_p$ (conjugated\cite{bs2} to $\Delta_{pd}$) between the $x\leq x_{cs}$ and $x\geq x_{cs}$ regimes are sizeable, $\varepsilon_d$ and $\varepsilon_p$ also vary due to the concomittant variation\cite{tu1,ku1} of the Madelung field. The variation of $\Delta_{pd}$ in the N\' eel (N) phase with respect to the value of $\Delta_{pd}^{(1)}$ for $x=0^+$ is

\begin{equation}
\delta\Delta_{pd}^M=\alpha_MV_{pd}(n_d^{(N)}-n_d^{(2)})\;,
\end{equation}

\noindent while $\bar\varepsilon$ is essentially unaffected for $x_{cs}=0^+$. The Madelung constant $\alpha_M$ is dominated\cite{tu1,ku1} by the first and second neighbor interactions $V_{pd}$ and $V_{pp}$, $\alpha_M\approx 2.14$. With repulsive\cite{em1} $V_{pd}\approx2$ eV and $\delta n_d\approx0.1-0.2$ the difference of excitation energies is thus increased further already in the iterative (rather than selfconsistent\cite{tu1}) Madelung scheme. Those observations reconcile, for $x\geq x_{cs}$, the ARPES data where the hole is created at the Fermi level and the measurement of the soft-X-ray absorption\cite{ch1} where the hole is annihilated at the Fermi level. In both cases, the Fermi level is consistently assumed to lie in the renormalized but still broad $L$-band of Fig.~\ref{fig004}. 

\subsection{Relation to superconductivity}

In our $U_d\rightarrow\infty$ limit the superconductivity is itself a coherent, translationally and locally gauge invariant\cite{an1} (double occupancy of Cu forbidden) state. It is therefore also expected to smoothen out the intrinsic disorders to some extent, but apparently it competes with the SDW and O$_x$/O$_y$ CT coherences. Extraordinarily, the superconductivity disappears entirely in the commensurate LTT phase of LBCO for $x\approx x_{vH}\approx1/8$ ($\mu\approx\omega_{vH}$) when the $\vec q=0$ O$_x$/O$_y$ CT gets frozen by its coupling to the commensurate $\pi[1,0]/a$ LTT$/(\varepsilon_{xx}-\varepsilon_{yy})$ lattice deformation accompanied\cite{ft1,tr3} by its $\vec q_0^{coll}$ satellites. This provides a striking evidence for the importance in high-T$_c$ superconductivity of D$_4$ symmetry (approximate in YBCO$_7$) involving the in plane oxygens. Actually, the superconductivity may even be enhanced when the B$_{1g}$ O$_x$/O$_y$ CT corresponds to the dynamic D$_4$ symmetric (intra- or inter-band) quadrupolar exciton since the latter can contribute to the bosonic glue.\cite{an1,zu1,bs2} However, as already mentioned, a coherent exciton is not observed within the frequency range accessible to B$_{1g}$ Raman scattering.\cite{si2} In contrast, the 340 cm$^{-1}$ B$_{1g}$ phonon in YBCO coupled to superconducting order is well known\cite{ts2,si2} for a long time.\cite{mf1,bs2} Notably in this respect, the superconductivity is stronger in YBCO than in the lanthanum cuprates, where the zone boundary $\pi[1,0]/a$ LTT$/(\varepsilon_{xx}-\varepsilon_{yy})$ phonons are evidently much heavier. Such issues apparently require further in-depth investigations, including the possibility of superconducting (rather than AF) space coherence among the RVB singlets.\cite{an1} At present we can only infer directly from Fig.~\ref{fig007} that the effective repulsion $U_{d\mu}=4t_{pd}^4\Lambda_0$ of Eq.~(\ref{Eq014}) may generate the Coulomb pseudo potential $\mu^*$ in the $pdp$-$pdp$ Cooper ladder channel,\cite{an1} which is not prohibitive to high-T$_c$ superconductivity, especially in materials like YBCO, where $n_d$  is smaller\cite{ku1} than in lanthanum cuprates.

\section{Summary}

\begin{figure}[b]

\begin{center}
{\scalebox{0.6}{\includegraphics{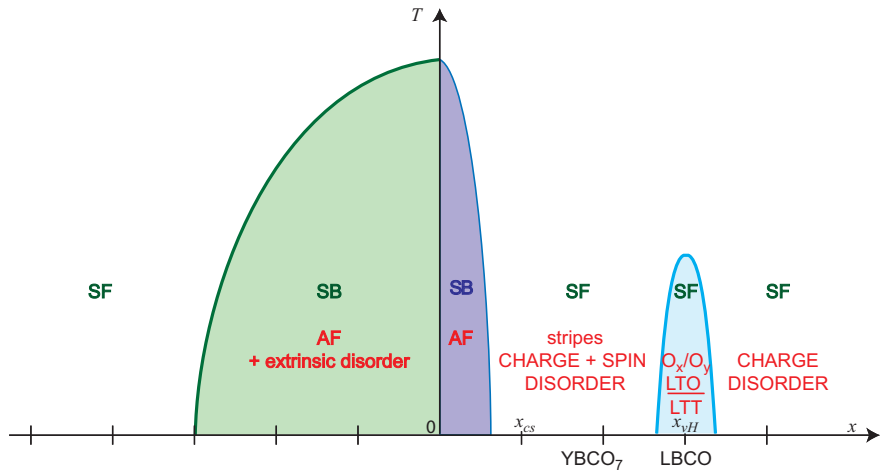}}}
\end{center}

\caption{(Color online) Cuprate behaviors associated to slave particle theories; SF stands for MSFT and SB for the anologous slave boson perturbation theory; shown are stripes, commensurate LTO/LTT instability and the spin plus charge disorder of the d$^{10}$$\leftrightarrow$d$^9$ type.\label{fig012}}

\end{figure}

Careful consideration of the overall experimental situation indicates the existence of a crossover between the insulating and metallic state for $x=x_{cs}(\Delta_{pd},t_{pd},t_{pp})$, which is 
quite small in actual cuprates since the model parameters obey $\Delta_{pd}^2\gtrsim2t_{pd}^2\gtrsim\Delta_{pd}|t_{pp}|\gtrsim4t_{pp}^2$.
The $x<x_{cs}$ regime can be associated with the Mott-AF or short range AF regime, while the $x>x_{cs}$ regime corresponds to the modified band picture with magnetic correlations. Such a distinction explains the persistence of magnetic correlations through the doping $x_{cs}$ and contrasts with the description of the hole doped metallic regime by the rigid $t-J$ model for all x>0 of interest. It is self-evident that the modified band regime also applies to the metallic state for $x<0$ (the MSFT is exact for $x=-1$), where the enhanced stability of the insulating AF phase is related\cite{xi1} to the pinning of doped electrons to the dopant sites (to the extrinsic disorder). Once the $x>x_{cs}$ time-dependent perturbation theory is shown to posses the local gauge invariance asymptotically, it is employed in the usual way, namely by summing up selectively chosen subseries, which emphasize the physically important features. The question left open here, whether or not the omission of the triplet $pdp$-$pdp$ scattering is sufficient to antisymmetrize the SFT {\it a posteriori}, thus becomes of limited practical importance.

This is summarized in Fig.~\ref{fig012}, which also shows the main physical features observed in cuprates, at least briefly mentioned here, namely the intrinsic CT d$^{10}$$\leftrightarrow$d$^9$ disorder, the stripes and the commensurate LTO/LTT instability driven by the coherent, purely intracell O$_x$/O$_y$ CT. In the metallic $x>x_{cs}$ regime all these are well described by the present theory, together with many other properties. Remarkably, the optimal doping in highest-T$_c$ cuprates, represented in Fig.~\ref{fig012} by YBCO$_7$, falls in the range $x_{cs}<x<x_{vH}$, while in lanthanates it is close to $x=x_{vH}$ where, moreover, superconductivity is suppressed\cite{ax1} by the commensurate LTO/LTT lattice instability.

In conclusion, the MSFT contains features which qualify it for further development in the context of the high T$_c$ superconductivity, just as do the corresponding moderate to strong coupling multiband theories in terms of finite $U_d$, some of which were discussed herein.

\begin{acknowledgments}

Many valuable discussions and correspondences with J. Friedel and L. P. Gor'kov are gratefully acknowledged. The authors are indebted to I. Kup\v ci\' c, D. K. Sunko and E. Tuti\v s for continuous collaboration, for critical reading of the manuscript and for important remarks. This work was supported by the Croatian Government under Projects $119-1191458-0512$ and $035-0000000-3187$.

\end{acknowledgments}

\appendix*

\section{Evaluation of \texorpdfstring{$B_\lambda^{(1)}$}{B}, \texorpdfstring{$F_\lambda^{(1)}$}{F} and \texorpdfstring{$\GD^{(1)}$}{Sigma}}

The purpose of this Appendix is to calculate explicitly $B_\lambda^{(1)}$, $F_\lambda^{(1)}$ and $\GD^{(1)}\sim B_\lambda^{(1)}\ast F_\lambda^{(1)}$ defined in Eq.~(\ref{Eq011}). $B_\lambda^{(1)}$ is given by

\begin{equation}
B_\lambda^{(1)}(\omega)=\frac{1}{\omega-\varepsilon_d-\lambda+i\eta-\beta_\lambda^{(1)}(\omega)}=B_0^{(1)}(\omega-\lambda),\label{A1}
\end{equation}

\noindent where $\beta_\lambda^{(1)}(\omega)$ is defined by Eq.~(\ref{Eq009}). Here $+i\eta$ keeps the memory of the pole in $B_\lambda^{(0)}$ while the poles in  $\beta_\lambda^{(1)}(\omega)$ are on the opposite side of the $\omega$-axis.  In the usual procedure, one uses the well known equality

\begin{equation}
\frac{1}{x\mp i\eta}=P\frac{1}{x}\pm i\pi\delta(x),\label{A2}
\end{equation}

\noindent and the $N\rightarrow\infty$ limit to calculate $\beta^{(1)}_\lambda(\omega)$ in integral form. It is then important to retain $+i\eta$ in Eq.~(\ref{A1}) because Eqs.~(\ref{Eq009}) and (\ref{A2}) show that $\IM\beta_\lambda^{(1)}(\omega)$ vanishes strictly everywhere except on a finite segment of the $\omega$-axis. $\IM B_\lambda^{(1)}$ turns out to be negative on the whole $\RE\omega$-axis as appropriate for bosons. The information about the side of the $\omega$-plane taken by the poles is, however, lost in this way. Analogously, $\IM F_\lambda^{(1)}$ changes sign twice which is far from the conventional (Fermi liquid) behavior of the fermion propagator. Additional care in the evaluation of $n_b^{(1)}$, $n_f^{(1)}$, $\tilde\varepsilon_d$, $A_{\vec k,\vec k'}$ in Eqs.~(\ref{Eq012a}) and (\ref{Eq012b}) is thus required.

To this end we consider Eq.~(\ref{A1}) as an equation involving a discrete set of poles, however dense. According to Eq.~(\ref{Eq009}) the propagator $B_\lambda^{(1)}$ contains the group of $N/2(1+x)$ (receding) poles in the upper half plane and one (advancing) pole in the lower half plane.  This means that $B_\lambda^(1)$ can be written in the form 

\begin{eqnarray}
B_\lambda^{(1)}&=&\frac{1+\tilde n_{\lambda b}^{(1)}(\omega)/2}{\omega-\tilde\varepsilon_{db}-\lambda+i\eta}+B_\lambda^{(1)<}\nonumber\\
B_\lambda^{(1)<}&=&-\frac{1}{N}\sum_{\vec k}\frac{\gamma_{\vec k}^{(\lambda b)}(\omega)f_{\vec k}^{(L)}}{\omega-\tilde\omega_b^{(L)}(\vec k)-\lambda+2i\eta}
\label{A3}
\end{eqnarray}                     

\noindent where the unique advancing pole $B_\lambda^{(1)>}$  is separated from $B_\lambda^{(1)<}$. Its position $\tilde\varepsilon_{db}+\lambda$ is shifted from $\varepsilon_d+\lambda$ according to $\varepsilon_{db}^{(1)}=\varepsilon_d+\beta_\lambda^{(1)}(\omega=\varepsilon_{db}^{(1)}+\lambda)$ , and, expanding $\beta_\lambda^{(1)}(\omega)$ around $\varepsilon_{db}^{(1)}+\lambda$ it is supplied with the spectral function $1+\tilde n_{\lambda b}^{(1)}(\omega)/2$ written in the way convenient for the boson on the $t>0$ side. We keep $\lambda$ in order to check how it evolves through various steps of the actual calculations. $B_\lambda^{(1)<}$ is represented as the sum of the receding poles shifted to the positions $\tilde\omega_b^{(L)}=\omega^{(L)}(\vec k)+\Delta\omega_b^{(L)}(\vec k)$ with the appropriate spectral densities $\gamma_{\vec k}^{(\lambda b)}(\omega)$.

It is, however, out of reach to evaluate analytically the energy shifts and the spectral functions for each of the $1+N(1+x)/2$ poles in Eq.~(\ref{A3}) for finite $N$ and therefore we turn to the $N\rightarrow\infty$ limit of Eq.~(\ref{A3}). One starts by using Eq.~(\ref{A2}) in Eq.~(\ref{Eq009}) for $\beta_\lambda^{(1)}(\omega)$ 

\begin{equation}
\IM \beta_\lambda^{(1)}(\omega)=-\frac{\pi t_{pd}^2}{N}\sum_{\vec k}z_{\vec k}^{(L)}f_{\vec k}^{(L)}\delta(\omega-\omega_{\vec k}^{(L)}-\lambda)\;.\label{A4}
\end{equation}

\noindent It is worthy of note that Eq.~(\ref{A4}) exhibits the unshifted positions and the spectral weights of the $N(1+x)/2$ poles. Those poles are grouped in the energy interval $[\omega_M^\lambda,\omega_\mu^\lambda]$ where $\omega_m^\lambda=\omega_M+\lambda$ and $\omega_\mu^\lambda=\mu^{(1)}+\lambda$ are the energies of the lowest and the highest occupied pole in the $L$-band shifted by $\lambda$. $\omega_M$ is the shorthand notation for the $\vec k=[\pi, \pi]$ energy in the $L$-band and $\mu^{(1)}$ is the HF chemical potential. 

We assert next that in the $N\rightarrow\infty$ limit

\begin{eqnarray}
\frac{1}{\pi}&&\frac{\IM \beta_\lambda^{(1)}(\omega)}{\omega-\varepsilon_d-\lambda-(\RE\beta_\lambda^{(1)}(\omega))^2+(\IM\beta_\lambda^{(1)}(\omega))^2}\nonumber\\&&\nonumber\\&&=\IM B_\lambda^{(1)<}(\omega)\label{A5}
\end{eqnarray}

\noindent using the expression (\ref{A4}) for $\IM\beta_\lambda^{(1)}(\omega)$. To show this we compare Eq.~(\ref{A5}) with Eq.~(\ref{A3}). All sums for $\IM\beta_\lambda^{(1)}(-\omega)$'s in Eq.~(\ref{A5}) should not be converted into integrals simultaneously. Neither should they all be kept as sums since the $\delta$-singularities should not appear simultaneously in the numerator and denominator of Eq.~(\ref{A5}). Indeed, according to Eqs.~(\ref{A3}) and (\ref{A2}) $\IM B_\lambda^{(1)<}$ can be expressed as the sum of $N(1+x)/2$ $\delta$-functions, associated with the values of $\omega^{(L)}(\vec k)$ shifted to $\tilde\omega^{(L)}(\vec k)=\omega^{(L)}(\vec k)+\Delta\omega^{(L)}_b(\vec k)$,

\begin{equation}
\IM B_\lambda^{(1)<}(\omega)=\frac{1}{N}\sum_{\vec k}\gamma_{\vec k}^{(\lambda b)}(\omega)f_{\vec k}^{(L)}\delta(\omega-\tilde\omega_b^{(L)}(\vec k)-\lambda)\;.\label{A6}
\end{equation}

\noindent The shift $\Delta\omega^{(L)}_b(\vec k)$ is of the order of $1/N$ in the $N\rightarrow\infty$ limit but, nevertheless, it has to be taken into account when defining this limit in Eq.~(\ref{A5}). This is most simply achieved by performing the $\vec k$-integration for $\RE\beta_\lambda^{(1)}$ and $\IM \beta_\lambda^{(1)}$ in the denominator of Eq.~(\ref{A5}) while still keeping $\IM \beta_\lambda^{(1)}$ in the numerator as a sum over $N(1+x)/2$ poles at the positions infinitesimally shifted to $\tilde\omega^{(L)}(\vec k)$. Such a procedure eliminates "on average" the $\delta$-functions from the denominator and leaves the infinitesimally shifted $\delta$-functions under the sum in the numerator.

The comparison of Eq.~(\ref{A6}) with such a modified Eq.~(\ref{A5}) allows us to identify $\gamma_{\vec k}^{(\lambda b)}(\omega)$ of Eq.~(\ref{A7}) as

\begin{equation}
\gamma_{\vec k}^{(\lambda b)}(\omega)=\frac{t_{pd}^2z_{\vec k}^{(L)}}{\omega-\varepsilon_d-\lambda-(\RE\beta_\lambda^{(1)}(\omega))^2+(\IM\beta_\lambda^{(1)}(\omega))^2}\;,\label{A7}
\end{equation}

\noindent for each of the $N(1+x)/2$ poles after letting $\Delta\omega_b^{(L)}(\vec k)\rightarrow0$. Whenever associated with $\delta(\omega-\tilde\omega_{\vec k}^\lambda)$ in its numerator $\gamma_{\vec k}^{(\lambda b)}(\omega)$ can be replaced by $\gamma_{\vec k}^{(\lambda b)}(\omega_{\vec k}^\lambda)$ by setting $\omega=\omega_{\vec k}^\lambda=\omega_{\vec k}^{(L)}+\lambda$ in the "$N$-averaged" denominator of Eq.~(\ref{A5}). It is obvious that such an $\IM B_\lambda^{(1)<}$ vanishes strictly outside the interval $[\omega_M^\lambda,\omega_\mu^\lambda]$ due to the fact that the $1/N$ shifts $\Delta\omega_b^{(L)}(\vec k)$ are neglected after the $N\rightarrow\infty$ limiting procedure. Importantly, $\RE\beta_\lambda^{(1)}$ is needed in Eq.~(\ref{A7}) in addition to $\IM\beta_\lambda^{(1)}$ to take care of the "average" spectral density $\gamma_{\vec k}^{(\lambda b)}$ of the "unshifted" receding poles. Inserting $\gamma_{\vec k}^{(\lambda b)}(\omega)$ of Eq.~(\ref{A7}) into Eq.~(\ref{A3}) now determines $B_\lambda^{(1)<}$.

Obviously, $\tilde n_{\lambda b}^{(1)}(\omega=\tilde\varepsilon_{db}+\lambda)$ in $B_\lambda^{(1)>}$ is nothing but $n_{b-}^{(1)}$, independent of $\lambda$. The structure of $n_b^{(1)}$ is however more transparent when calculated on the $t<0$ side from $n_{b+}^{(1)}$,  

\begin{equation}
n_{b+}^{(1)}=-2\int_{\omega_M^\lambda}^{\omega_\mu^\lambda}\IM B_\lambda^{(1)<}(\omega)d\omega\;.\label{A8}
\end{equation}

\noindent Eq.~(\ref{A8}) can be evaluated in two ways. One can insert the result (\ref{A7}) for $\gamma_{\vec k}^{(\lambda b)}(\omega)$ in Eq.~(\ref{A8}), carry out the $\omega$-integration and then the $\vec k$-integration. Alternatively one can transform the numerator of Eq.~(\ref{A5}) immediately in the integral over $\vec k$ and then carry out the $\omega$-integration in Eq.~(\ref{A8}). The result is the same, i.e. the two integrations in question are interchangeable in the $N\rightarrow\infty$ limit. Noting that $\Delta n_b^{(1)}$ of Eq.~(\ref{Eq010a}) corresponds to omitting $\beta_\lambda^{(1)}(\omega)$ in the denominator of $\gamma_{\vec k}^{(\lambda b)}(\omega)$, it is immediately apparent that $n_b^{(1)}\approx\Delta n_b^{(1)}=n_d^{(1)}$, provided that $n_d^{(1)}$ is small.

The analogous $b\leftrightarrow f$ symmetric procedure can be carried out for the spinless fermion $F_\lambda^{(1)}(\omega)=F_0^{(1)}(\omega-\lambda)$, as already mentioned in connection with Eq.~(\ref{Eq010b}). In $\Delta Q_{\vec R}^{(1)}$ associated with $\Delta B_\lambda^{(1)}$ and $\Delta F_\lambda^{(1)}$ of Figs.~\ref{fig003a} and \ref{fig003b} one sums over two bosons according to Eq.~(\ref{A8}), and $\Delta Q_{\vec R}^{(1)} =0$. However, $Q_{\vec R}^{(1)}$ differs from unity due to the asymmetry of the roles of the spin factor 2 in $B_\lambda^{(1)}$ and $F_\lambda^{(1)}$. In particular the receding pole in $F_\lambda^{(1)}$ shifts from $\lambda$ to $\lambda-2\beta_\lambda^{(1)}(\omega=\varepsilon_{df}^{(1)}+\lambda)$. Similarly, $\gamma_{\vec k}^{(\lambda f)}(\omega)$ differs from $\gamma_{\vec k}^{(\lambda b)}(\omega)$ of Eq.~(\ref{A7}) by the appearance of the factor 2 in the denominator. It is then clear that $Q_{\vec R}^{(1)}\approx1$ for small $n_d^{(1)}$, when $n_b^{(1)}\approx\Delta n_b^{(1)}$ and $n_f^{(1)}\approx\Delta_f^{(1)}$. 

Finally, using $B_\lambda^{(1)}$ of Eq.~(\ref{A3}) and its $b\leftrightarrow f$ symmetric counterpart $F_\lambda^{(1)}$ one finds that 

\begin{equation}
\varepsilon_d^{(1)}=\varepsilon_d+3\beta_\lambda^{(1)}(\omega=\varepsilon_d^{(1)}+\lambda)\label{A9}
\end{equation}

\begin{widetext}

\begin{equation}
\GD^{(1)<}=\frac{1}{N^2}\sum_{\vec k',\vec k''}f^{(L)}_{\vec k'}f_{\vec k''}^{(L)}\frac{\gamma_{\vec k'}^{(\lambda b)}(\hat\omega_1)\gamma_{\vec k''}^{(\lambda f)}(\hat\omega_1-\omega)+\gamma_{\vec k'}^{(\lambda b)}(\hat\omega_2)\gamma_{\vec k''}^{(\lambda f)}(\hat\omega_2-\omega)}{\omega-\omega_{\vec k'}^{(L)}-\omega_{\vec k''}^{(L)}+\varepsilon_d-4i\eta}\;.\label{A10}
\end{equation}

\end{widetext}

\noindent $\varepsilon_d^{(1)}$ appears in $\GD^{(1)>}$ of Eq.~(\ref{Eq012a}) which corresponds to the convolution of only two $b\leftrightarrow f$ symmetric poles $B_\lambda^{(1)>}$ and $F_\lambda^{(1)<}$. $\GD^{(1)<}$ is taken here in the $N\rightarrow\infty$ limit with the short hand notations $\hat\omega_1=\omega_{\vec k}^{(L)}+\lambda$ and $\hat\omega_2=\omega-\varepsilon_{\vec k'}^{(L)}+\varepsilon_d+\lambda$. As easily seen $\GD^{(1)<}$ is, like $\GD^{(1)>}$, independent of  $\lambda$. Equation~(\ref{A10}) provides the closed $N\rightarrow\infty$ expression for $A_{\vec k',\vec k''}$ in Eq.~(\ref{Eq012b}). It can be easily turned into integration by using Eq.~(\ref{A2}).


\begin{thebibliography}{99}


\bibitem{fr1} J. Friedel,
	J. Phys. Cond. Matt. {\bf 1}, 7757 (1989).

\bibitem{an1} P. W. Anderson,
	Science {\bf 317}, 1705 (2007).

\bibitem{in1} A. Ino {\it et al.},
	Phys. Rev. B {\bf 65}, 094504 (2002).

\bibitem{zu1} X. J. Zhou, T. Cuk, T. Devereaux, N. Nagaosa, and Z.-X. Shen,
	Handbook of High-Temperature Supercoductivity,  p. 87 (ed.  J. R. Schrieffer,  J. S. Brooks, Springer Verlag, 2007).

\bibitem{ku1} I. Kup\v ci\' c, S. Bari\v si\' c, and E. Tuti\v s,
	Phys. Rev. B {\bf 57}, 8590 (1998).

\bibitem{hr1} M. Horvati\' c, P. Butaud, P. S\' egransan, Y. Berthier, C. Berthier, J. Y. Henry, and M. Couach,
	Physica C {\bf 159}, 68 (1989).

\bibitem{tg1} M. Takigawa, A. P. Reyes, P. C. Hammel, J. D. Thompson, R. H. Heffner, Z. Fisk, and K. C. Ott,
	Phys. Rev. B {\bf 43}, 247 (1991).

\bibitem{mt1} R. L. Martin,
	Phys. Rev. Lett. {\bf 75}, 744 (1995).

\bibitem{sz1} M. Suzuki,
	Phys. Rev. B {\bf 39}, 2312 (1989).

\bibitem{wa1} Y. Watanabe, Z. Z. Wang, S. A. Lyon, D. C. Tsui, N. P. Ong, J. M. Tarascon, and P. Barboux,
	Phys. Rev. B {\bf 40}, 6884 (1989).

\bibitem{bo1} G. S. Boebinger {\it et al.},
	Phys. Rev. Lett. {\bf 77}, 5417 (1996).

\bibitem{ma1} A. Malinowski {\it et al.},
	Phys. Rev. Lett. {\bf 79}, 495 (1997).

\bibitem{uc1} S. Uchida, T. Ido, H. Takagi, T. Arima, Y. Tokura, and S. Tajima,
	Phys. Rev. B {\bf 43}, 7942 (1991).

\bibitem{gz1} A. G. Gozar, S. Komiya, Y. Ando and G. Blumberg,
	Frontiers in Magn. Materials, 755, (Ed. A.V.Narlikar, Springer-Verlag, 2005).

\bibitem{ts1} L. Tassini, W. Prestel, A. Erb, M. Lambacher, and R. Hackl, 
	Phys. Rev. B {\bf 78}, 020511(R) (2008).

\bibitem{si2} S. Sugai, H. Suzuki, Y. Takayanagi, T. Hosokawa, and N. Hayamizu,
      Phys. Rev. B {\bf 68}, 184504 (2003).
\bibitem{bl1} G. Blumberg, P. Abbamonte, M. V. Klein, W. C. Lee, D. M. Ginsberg, L. L. Miller, and A. Zibold,
	Phys. Rev. B {\bf 53}, R11930 (1996).

\bibitem{tr2} J. M. Tranquada {\it et al.},
	Nature {\bf 429}, 534 (2004).

\bibitem{tr3} J. M. Tranquada {\it et al.},
	Phys. Rev. B {\bf 78}, 174529 (2008).

\bibitem{st1} C. Stock {\it et al.},
	Phys. Rev B {\bf 69}, 014502 (2004);
	{\bf 71}, 024522 (2005).

\bibitem{ki1} S. A. Kivelson, I. P. Bindloss, E. Fradkin, V. Oganesyan, J. M. Tranquada, A. Kapitulnik, and C. Howald,
	Rev. Mod. Phys. {\bf 75}, 1201 (2003).

\bibitem{ya1} K. Yamada {\it et al.},
	Phys. Rev. B {\bf 57}, 6165 (1998).

\bibitem{vt1} M. Vojta,
	Adv. Phys. {\bf 58}, 564 (2009).

\bibitem{bi1} A. Bianconi {\it et al.}, 
	Phys. Rev. Lett. {\bf 76}, 3412 (1996).

\bibitem{sg1} S. Sugita, T. Watanabe, and A. Matsuda,
	Phys. Rev. B {\bf 62}, 8715 (2000).

\bibitem{bb1} S. Bari\v si\' c and O. S. Bari\v si\' c,
	Physica B {\bf 404}, 370 (2009).

\bibitem{ax1} J. D. Axe, A. H. Moudden, D. Hohlwein, D. E. Cox, K. M. Mohanty, A. R. Moodenbaugh, and Youwen Xu,
	Phys. Rev. Lett. {\bf 62}, 2751 (1989).

\bibitem{kmr} H. Kimura, Y. Noda, H. Goka, M. Fujita, K. Yamada, and G. Shirane,
	J. Phys. Soc. Jpn. {\bf 74}, 445 (2005).

\bibitem{ft1} M. Fujita, H. Goka, K. Yamada, J. M. Tranquada, and L. P. Regnault,
	Phys. Rev. B {\bf 70}, 104517 (2004).

\bibitem{yo1} T. Yoshida {\it et al.}, 
	Phys. Rev. B {\bf 74}, 224510 (2006).

\bibitem{vl1} T. Valla, A. V. Fedorov, J. Lee, J. C. Davis, and G. D. Gu, 	Science {\bf 314}, 1914 (2006). 

\bibitem{al1} A. Alloul, J. Bobrof, M. Gabay, and P. J. Hirschfeld,
	Rev. Mod. Phys. {\bf 81}, 45 (2009).

\bibitem{go3} L. P. Gor'kov and G. B. Teitelbaum,
	J. Phys.: Conf. Ser. {\bf 108}, 012009 (2008).

\bibitem{jl1} M.-H. Julien {\it et al.},
	Phys. Rev. B {\bf 63}, 144508 (2001).

\bibitem{cp1} S. L. Cooper, F. Slakey, M. V. Klein, J. P. Rice, E. D. Bukowski, and D. M. Ginsberg,
	Phys. Rev. B {\bf 38}, 11934 (1988).

\bibitem{em1} V. J. Emery,
	Phys. Rev. Lett. {\bf 58}, 2794 (1987).

\bibitem{km1} Ju H. Kim, K. Levin, and A. Auerbach,
	Phys. Rev. B {\bf 39}, 11633 (1989).

\bibitem{bs5} S. Bari\v si\' c and I. Batisti\' c,
	Pysica Scripta. {\bf T27}, 78 (1989).

\bibitem{fr3} J. Friedel and M.Kohmoto,
	Eur. Phys. J. B {\bf 30}, 427 (2002).

\bibitem{ka1} J. Kanamori,
	Prog. Theor. Phys. {\bf 30}, 275 (1963).

\bibitem{dz2} I. E. Dzyaloshinskii and V. M. Yakovenko,
	ZhETF 94, {\bf 344} (1988).

\bibitem{ru1} J. Ruvalds and A. Virosztek,
	Phys. Rev. B {\bf 43}, 5498 (1991).

\bibitem{bs2} S. Bari\v si\' c,
	Int. Journ. Mod. Phys. B {\bf 5}, 2439 (1991).

\bibitem{ku3} I. Kup\v ci\' c and S. Bari\v si\' c,
	Phys. Rev. B {\bf 75}, 094508 (2007). 

\bibitem{ku2}  I. Kup\v ci\' c,
	Physica C {\bf 391}, 251 (2003).

\bibitem{pu1} B. Piveteau and C. Nougera,
       Fizika {\bf 21}, 237 (1989).

\bibitem{sc1} H. J. Schulz,
	 Phys. Rev. Lett. {\bf 64}, 1445 (1990).

\bibitem{sk3} D. K. Sunko {\it et al.},
	(unpublished).

\bibitem{mr1} I. Mrkonji\' c and S. Bari\v si\' c,
	Eur. Phys. J. B {\bf 34}, 69 (2003);
	{\bf 34}, 441 (2003).

\bibitem{xu1} J. X. Xu, T. J. Watson-Yang, J. Yu, and A. J. Freeman,
	Phys. Lett. A {\bf 120}, 489 (1987).  

\bibitem{qs1} Q. Si, Y. Zha, K. Levin, and J. P. Lu,
	Phys. Rev. B {\bf 47}, 9055 (1993).

\bibitem{go1} L. P. Gor'kov and A. V. Sokol,
	JETP Lett. {\bf 46}, 420 (1987).

\bibitem{ni1} H. Nik\v si\' c, E. Tuti\v s, and S. Bari\v si\' c,
	Physica C {\bf 241}, 247 (1995).

\bibitem{zl1} M. B. Z\" olfl, Th. Maier, Th. Pruschke, and J. Keller,
	Eur. Phys. B {\bf 13}, 47 (2000).

\bibitem{mc1} A. Macridin, M. Jarrell, Th. Maier, and G. A. Sawatzky,
	Phys. Rev. B {\bf 71}, 134527 (2005).

\bibitem{rb1} A. N. Rubtsov, M. I. Katsnelson, A. I. Lichtenstein, and A. Georges,
	Phys. Rev. B {\bf 79}, 045133 (2009).

\bibitem{lo1} J. Lorenzana and G. A. Sawatzky,
	Phys. Rev. B {\bf 52}, 9576 (1995).

\bibitem{ko1} B. G. Kotliar, P. A. Lee, and N. Read,
	Physica C {\bf 153}, 538 (1988).

\bibitem{tu1} E. Tuti\v s,
	{\it Ph.D. Thesis} (University of Zagreb, 1994);
	E. Tuti\v s, H. Nik\v si\' c, and S. Bari\v si\' c,
	Lect. Notes Phys. {\bf 427}, 1 (1997). 

\bibitem{sk2} D. K. Sunko and S. Bari\v si\' c,
	Phys. Rev. B {\bf 75}, 060506(R) (2007).

\bibitem{bbk} S. Bari\v si\' c, O. S. Bari\v si\' c, and I. Kup\v ci\' c,
	(unpublished).  

\bibitem{li1} Y. M. Li, D. N. Sheng, Z. B. Su, and L. Yu,
	Phys. Rev. B {\bf 45}, 5428 (1992).

\bibitem{za1} F. C. Zhang and T. M. Rice,
	Phys. Rev. B {\bf 37}, 3759 (1988).

\bibitem{xi1} T. Xiang, H. G. Luo, D. H. Lu, K. M. Shen, and Z. X. Shen,
	Phys. Rev. B {\bf 79}, 014524 (2009).

\bibitem{ch1} C. T. Chen {\it et al.},
	Phys. Rev. Lett. {\bf 66}, 104 (1991).

\bibitem{ge1} W. Geertsma,
	{\it Ph.D. thesis} (Univ. of Groningen, 1979);
	Physica {\bf B164}, 241 (1990).

\bibitem{fu1} S. Fukagawa, A. Kobayashi, K. Miura, T. Matsuura, and Y. Kuroda, 
	J. Phys. Soc. Jpn. {\bf 67}, 3536 (1998).

\bibitem{sk4} D. K. Sunko,
	JETP {\bf 136}, 758 (2009).

\bibitem{cl1} R. Coldea, S. M. Hayden, G. Aeppli, T. G. Perring, C. D. Frost, T. E. Mason, S.-W. Cheong, and Z. Fisk,
	Phys. Rev. Lett. {\bf 86}, 5377 (2001). 

\bibitem{wl1} A. C. Walters, T. G. Perring, J.-S. Caux, A. T. Savici, G. D. Gu, C.-C. Lee, W. Ku, and I. A. Zaliznyak,
	Nature Phys. {\bf 5}, 867 (2009).

\bibitem{pe1} A. G. Petukhov, I. I. Mazin, L. Chioncel, and A. I. Lichtenstein,
	Phys. Rev. B {\bf 67}, 153106 (2003).

\bibitem{bn1} S. E. Barnes,
	J. Phys. F {\bf 6}, 1375 (1976). 

\bibitem{co1} P. Coleman,
	Phys. Rev. B {\bf 29}, 3035 (1984).

\bibitem{ho1} A. Houghton, N. Read, and H. Won,
	Phys. Rev. B {\bf 37}, 3782 (1988).

\bibitem{kr1} J. Kroha and P. W\" olfle,
	J. Phys. Soc. Jpn. {\bf 74}, 16 (2005).

\bibitem{an2} P. W. Anderson,
	Adv. Phys. {\bf 46}, 3 (1997).

\bibitem{ob1} O. S. Bari\v si\' c, 
	Phys. Rev. B {\bf 76}, 193106 (2007).

\bibitem{msh} A. Messsiah,
	{\it Quantum Mechanics II}, ch. XIV (North-Holland Publ. Co., 1967).

\bibitem{yi1} Y.-J. Kao, Q. Si, and K. Levin,
	Phys. Rev. B {\bf 61}, R11898 (2000).

\bibitem{se1} S. E. Sebastian, N. Harrison, E. Palm, T. P. Murphy, C. H. Mielke, R. Liang, D. A. Bonn, W. N. Hardy, and G. G. Lonzarich1, 
	Nature {\bf 454}, 200 (2008).

\bibitem{sch} M. C. Schabel, C. H. Park, A. Matsuura, Z. X. Shen, D. A. Bonn, X. Liang, and W. N. Hardy, 
	Phys. Rev. B {\bf 57}, 6107 (1998).

\bibitem{rl1} R. Liu, B. W. Veal, A. P. Paulikas, J. W. Downey, P. J. Kosti\' c, S. Fleshler, U. Welp, C. G. Olson, X. Wu, A. J. Arko, and J. J. Joyce,
	Phys. Rev. B {\bf 46}, 11056 (1992).

\bibitem{mi1} K. Maiti, J. Fink, S. de Jong, M. Gorgoi, C. T. Lin, M. Raichle, V. Hinkov, M. Lambacher, A. Erb, and M. S. Golden,
	Phys. Rev. B {\bf 80}, 165132 (2009).

\bibitem{pnc} L. Pinchovius and M. Braden,
	Phys. Rev. B {\bf 60}, R15039 (1999).

\bibitem{lo2} J. Lorenzana and G. Seibold, 
	Phys. Rev. Lett. {\bf 89}, 136401 (2002). 
\bibitem{va2} C. M. Varma,
	Phys. Rev. B {\bf 73}, 155113 (2006).

\bibitem{si3} S. Sugai, Y. Takayanagi, and N. Hayamizu,
	Phys. Rev. Lett. {\bf 96}, 137003 (2006).

\bibitem{lu1} D. H. Lu {\it et al.},
	Phys. Rev. Lett. {\bf 86}, 4370 (2001).

\bibitem{kk1} A. Kaminski, M. Randeria, J. C. Campuzano, M. R. Norman, H. Fretwell, J. Mesot, T. Sato, T. Takahashi, and K. Kadowaki,
	Phys. Rev. Lett. {\bf 86}, 1070 (2001).

\bibitem{no1} M. R. Norman,
	{\it Handbook of Magnetism and Magnetic Materials} {\bf 5}, 2671 (ed. H. Kronmuller and S. Parkin, Willey, NY, 2007).

\bibitem{ts2} L. Tassini, F. Venturini, Q.-M. Zhang, R. Hackl, N. Kikugawa, and T. Fujita,
	Phys. Rev. Lett. {\bf 95}, 117002 (2005).

\bibitem{on1} S. Ono, S. Komiya, and Y. Ando,
	Phys. Rev. B {\bf 75}, 024515 (2007).

\bibitem{mf1} R. M. MacFarlane, H. Rosen, and H. Seki,
	Sol. St. Comm. {\bf 63}, 839 (1987).


\end{thebibliography}
\end{document}